\documentclass[english,reprint,longbibliography]{revtex4-1}
\usepackage[T1]{fontenc}
\usepackage[latin9]{inputenc}
\usepackage{color}
\usepackage{babel}
\usepackage{verbatim}
\usepackage{amsmath}
\usepackage{amssymb}
\usepackage{graphicx}
\usepackage{xargs}[2008/03/08]
\usepackage[unicode=true,pdfusetitle,
 bookmarks=true,bookmarksnumbered=true,bookmarksopen=true,bookmarksopenlevel=1,
 breaklinks=false,pdfborder={0 0 0},pdfborderstyle={},backref=false,colorlinks=true]
 {hyperref}

\makeatletter

\newcommand{\lyxdot}{.}

\makeatother

\begin{document}

\title{Optimal adaptive control for quantum metrology with time-dependent
Hamiltonians}

\author{Shengshi Pang$^{1,2}$}
\email{Correspondence and requests for materials should be addressed to Shengshi Pang (email: pangshengshi@gmail.com).}

\author{Andrew N. Jordan$^{1,2,3}$}

\affiliation{$^{1}$Department of Physics and Astronomy, University of Rochester,
Rochester, New York 14627, USA}

\affiliation{$^{2}$Center for Coherence and Quantum Optics, University of Rochester,
Rochester, New York 14627, USA}

\affiliation{$^{3}$Institute for Quantum Studies, Chapman University, 1 University
Drive, Orange, CA 92866, USA}
\begin{abstract}
Quantum metrology has been studied for a wide range of systems with
time-independent Hamiltonians. For systems with time-dependent Hamiltonians,
however, due to the complexity of dynamics, little has been known
about quantum metrology. Here we investigate quantum metrology with
time-dependent Hamiltonians to bridge this gap. We obtain the optimal
quantum Fisher information for parameters in time-dependent Hamiltonians,
and show proper Hamiltonian control is necessary to optimize the Fisher
information. We derive the optimal Hamiltonian control, which is generally
adaptive, and the measurement scheme to attain the optimal Fisher
information. In a minimal example of a qubit in a rotating magnetic
field, we find a surprising result that the fundamental limit of $T^{2}$
time scaling of quantum Fisher information can be broken with time-dependent
Hamiltonians, which reaches $T^{4}$ in estimating the rotation frequency
of the field. We conclude by considering level crossings in the derivatives
of the Hamiltonians, and point out additional control is necessary
for that case.
\end{abstract}
\maketitle
\newcommandx\hgt[2][usedefault, addprefix=\global, 1=t, 2=g]{H_{#2}(#1)}
\newcommandx\pg[1][usedefault, addprefix=\global, 1=g]{\partial_{#1}}
\newcommandx\phgt[2][usedefault, addprefix=\global, 1=t, 2=g]{\pg[#2]\hgt[#1][#2]}
\global\long\def\si{|\psi_{0}\rangle}
\global\long\def\st{|\psi_{T}\rangle}
\newcommandx\fq[1][usedefault, addprefix=\global, 1=g]{I_{#1}^{(Q)}}
\global\long\def\ig{I_{g}}
\global\long\def\csi{\langle\psi_{0}|}
\newcommandx\hg[2][usedefault, addprefix=\global, 1=g, 2=T]{h_{#1}(#2)}
\newcommandx\ut[2][usedefault, addprefix=\global, 1=0, 2=T]{U(#1\rightarrow#2)}
\newcommandx\utd[2][usedefault, addprefix=\global, 1=0, 2=T]{U^{\dagger}(#1\rightarrow#2)}
\newcommandx\utg[3][usedefault, addprefix=\global, 1=0, 2=g, 3=T]{U_{#2}(#1\rightarrow#3)}
\newcommandx\utgd[3][usedefault, addprefix=\global, 1=0, 2=g, 3=T]{U_{#2}^{\dagger}(#1\rightarrow#3)}
\global\long\def\i{i}
\newcommandx\lmax[1][usedefault, addprefix=\global, 1=T]{\lambda_{\max}(#1)}
\newcommandx\lmin[1][usedefault, addprefix=\global, 1=T]{\lambda_{\min}(#1)}
\newcommandx\lm[2][usedefault, addprefix=\global, 1=\hg, 2=\max/\min]{\lambda_{#2}(#1)}
\global\long\def\ket#1{|#1\rangle}
\newcommandx\bra[1][usedefault, addprefix=\global, 1=]{\langle#1|}
\global\long\def\e{e}
\global\long\def\d{{\rm d}}
\global\long\def\dg{\delta g}
\global\long\def\et{\e^{\i\theta}}
\newcommandx\klmax[2][usedefault, addprefix=\global, 1=t, 2=]{\ket{#2\psi_{\max}(#1)}}
\newcommandx\klmin[2][usedefault, addprefix=\global, 1=t, 2=]{\ket{#2\psi_{\min}(#1)}}
\newcommandx\klm[3][usedefault, addprefix=\global, 1=t, 2=\max/\min, 3=]{\ket{#3\psi_{#2}(#1)}}
\newcommandx\blm[3][usedefault, addprefix=\global, 1=t, 2=k, 3=]{\bra[#3\psi_{#2}(#1)]}
\global\long\def\delt{\Delta t}
\global\long\def\dt{\d t}
\global\long\def\ss{|\psi\rangle}
\global\long\def\css{\langle\psi|}
\newcommandx\hc[1][usedefault, addprefix=\global, 1=t]{H_{{\rm c}}(#1)}
\newcommandx\htot[1][usedefault, addprefix=\global, 1=t]{H_{{\rm tot}}(#1)}
\global\long\def\pt{\partial_{t}}
\global\long\def\ptt{\partial_{t}^{2}}
\newcommandx\thk[2][usedefault, addprefix=\global, 1=t, 2=k]{\theta_{#2}(#1)}
\newcommandx\thkd[2][usedefault, addprefix=\global, 1=t, 2=k]{\dot{\theta}_{#2}(#1)}
\newcommandx\etk[1][usedefault, addprefix=\global, 1=k]{\e^{\i\thk[][#1]}}
\newcommandx\netk[2][usedefault, addprefix=\global, 1=k, 2=t]{\e^{-\i\thk[#2][#1]}}
\newcommandx\fkt[2][usedefault, addprefix=\global, 1=t, 2=k]{f_{#2}(#1)}
\global\long\def\hgg{H_{g}}
\newcommandx\bt[1][usedefault, addprefix=\global, 1=t]{\mathbf{B}(#1)}
\global\long\def\sig{\boldsymbol{\sigma}}
\newcommandx\nt[1][usedefault, addprefix=\global, 1=t]{\mathbf{n}(#1)}
\newcommandx\tpk[1][usedefault, addprefix=\global, 1=]{\tau_{#1}^{+}}
\newcommandx\tmk[1][usedefault, addprefix=\global, 1=]{\tau_{#1}^{-}}
\global\long\def\tpmk{\tau^{\pm}}
\newcommandx\tk[1][usedefault, addprefix=\global, 1=]{\tau_{#1}}
\newcommandx\ha[1][usedefault, addprefix=\global, 1=t]{H_{a}(#1)}
\newcommandx\hkt[2][usedefault, addprefix=\global, 1=t, 2=]{h_{#2}(#1)}
\global\long\def\sdt{\delta t}
\newcommandx\kk[1][usedefault, addprefix=\global, 1=k]{\ket{#1}}
\newcommandx\bj[1][usedefault, addprefix=\global, 1=j]{\bra[#1]}
\global\long\def\hp{H^{\prime}(t)}
\global\long\def\smn{\sigma_{mn}}
\global\long\def\sx{\sigma_{X}}
\global\long\def\sy{\sigma_{Y}}
\global\long\def\sz{\sigma_{Z}}
\newcommandx\ham[1][usedefault, addprefix=\global, 1=t]{H(#1)}
\newcommandx\etkj[2][usedefault, addprefix=\global, 1=k, 2=j]{\e^{\i(\thk[][#1]-\thk[][#2])}}
\global\long\def\upt{U^{\prime}(t)}
\global\long\def\tp{t^{\prime}}
\global\long\def\dtp{\d\tp}
\global\long\def\pb{\partial_{B}}
\global\long\def\ko{\ket 0}
\global\long\def\kl{\ket 1}
\global\long\def\ex{\mathbf{e}_{x}}
\global\long\def\ey{\mathbf{e}_{y}}
\global\long\def\ez{\mathbf{e}_{z}}
\global\long\def\po{\partial_{\omega}}
\global\long\def\bo{B_{0}}
\global\long\def\mt{\mathcal{T}}
\global\long\def\sg{\psi_{g}}
\global\long\def\ksg{\ket{\psi_{g}}}
\global\long\def\bsg{\bra[\sg]}
\global\long\def\avg#1{\langle#1|#1\rangle}
\global\long\def\ovl#1#2{\langle#1|#2\rangle}
\global\long\def\oc{\omega_{{\rm c}}}
\global\long\def\oo{\omega_{0}}
\global\long\def\pkg{p_{g}(X)}
\global\long\def\eg{\hat{g}}
\global\long\def\ho{H_{0}}
\newcommandx\var[2][usedefault, addprefix=\global, 1=]{{\rm Var}[#2]_{#1}}
\newcommandx\ug[1][usedefault, addprefix=\global, 1=]{U_{g}^{#1}}
\global\long\def\h{h_{g}}
\global\long\def\ksgt{\ket{\sg(T)}}
\global\long\def\hot{\ho(t)}
\newcommandx\umax[1][usedefault, addprefix=\global, 1=t]{\mu_{\max}(#1)}
\newcommandx\umin[1][usedefault, addprefix=\global, 1=t]{\mu_{\min}(#1)}
\newcommandx\uk[2][usedefault, addprefix=\global, 1=t, 2=k]{\mu_{#2}(#1)}
\global\long\def\kp{\ket +}
\global\long\def\km{\ket -}
\global\long\def\kpm{\ket{\pm}}
\global\long\def\do{\delta\omega}
\global\long\def\sl{|\psi_{1}\rangle}
\global\long\def\gc{g_{{\rm c}}}
\newcommandx\ktlm[2][usedefault, addprefix=\global, 1=, 2=t]{|#1\widetilde{\psi_{k}}(#2)\rangle}
\global\long\def\btlm{\langle\widetilde{\psi_{k}}(t)|}
\newcommandx\ktlmax[2][usedefault, addprefix=\global, 1=, 2=t]{|#1\widetilde{\psi_{\max}}(#2)\rangle}
\newcommandx\ktlmin[2][usedefault, addprefix=\global, 1=, 2=t]{|#1\widetilde{\psi_{\min}}(#2)\rangle}
\global\long\def\poct{\partial_{\oc}^{2}}
\global\long\def\poc{\partial_{\oc}}
\global\long\def\av#1{\langle#1\rangle}
\newcommandx\bklm[5][usedefault, addprefix=\global, 1=k, 2=j, 3=t, 4=, 5=]{\langle#4\psi_{#1}(#3)|#5\psi_{#2}(#3)\rangle}
\global\long\def\vt{V(t)}
\global\long\def\vtd{V^{\dagger}(t)}
\global\long\def\pht{\ket{\phi(t)}}
\global\long\def\ppht{\ket{\pt\phi(t)}}
\global\long\def\vpht{\ket{\varphi(t)}}
\global\long\def\htp{H_{{\rm tot}}^{\prime}(t)}
\global\long\def\bp{\bra[+]}
\global\long\def\bmm{\bra[-]}
\global\long\def\mo{\mathcal{O}}
\newcommandx\pst[2][usedefault, addprefix=\global, 1=t, 2=]{\ket{\Psi_{#2}(#1)}}
\global\long\def\tpp{t^{\prime\prime}}

Precision measurement has been long pursued due to its vital importance
in physics and other sciences. Quantum mechanics supplies this task
with two new elements. On one hand, quantum mechanics imposes a fundamental
limitation on the precision of measurements, apart from any external
noise, the quantum noise \cite{Haus1962}, which is rooted in the
stochastic nature of quantum measurement and manifested by the Heisenberg
uncertainty principle. On the other hand, quantum mechanics also opens
new possibilities for improving measurement sensitivities by utilizing
non-classical resources, such as quantum entanglement and squeezing
\cite{Caves1981}. These have given rise to the wide interest in quantum
parameter estimation \cite{Helstrom1976,Holevo1982} and quantum metrology
\cite{Giovannetti2004,Giovannetti2011}. Since its birth, quantum
metrology has been applied in many areas, ranging from gravitational
wave detection \cite{Schnabel2010,Danilishin2012,Adhikari2014}, quantum
clocks \cite{Derevianko2011,Ludlow2015}, quantum imaging \cite{Kolobov1999,Lugiato2002,Dowling2015},
to optomechanics \cite{Aspelmeyer2014}, quantum biology \cite{Taylor2016},
etc. Various quantum correlations have been shown useful for enhancing
measurement sensitivities, including spin squeezed states \cite{Wineland1992,Ma2009,Hyllus2010,Gross2012,Rozema2014,Yukawa2014},
N00N states \cite{Lee2002,Resch2007,Nagata2007,Jones2009,Israel2014},
etc. Nonlinear interactions have been exploited to break the Heisenberg
limit even without entanglement \cite{Luis2004,Boixo2007,Roy2008,Boixo2008,Pezze2009,Napolitano2011,Hall2012a,Zwierz2014}.
For practical applications where disturbance from the environment
is inevitable, quantum metrology in open systems has been studied
\cite{Escher2011,Demkowicz-Dobrzanski2012,Chin2012,Tsang2013,Alipour2014},
and quantum error correction schemes for protecting quantum metrology
against noise have been proposed \cite{Tan2013,Arrad2014,Dur2014,Kessler2014,Lu2015a}.

While most previous research on quantum metrology was focused on multiplicative
parameters of Hamiltonians, growing attention has recently been drawn
to more general parameters of Hamiltonians \cite{Pang2014b} or physical
dynamics \cite{Liu2015,Jing2015}, such as those of magnetic fields
\cite{Pang2014b,Skotiniotis2015,Baumgratz2016,Yuan2016}. Interestingly,
in contrast to estimation of multiplicative parameters, estimation
of general Hamiltonian parameters exhibits distinct characteristics
in some aspects, particularly in the time scaling of the Fisher information
\cite{Pang2014b}, and often requires quantum control to gain the
highest sensitivity \cite{Yuan2015a}.

While there has been tremendous research devoted to quantum metrology,
most of those works were focused on time-independent Hamiltonians,
and little has been known when the Hamiltonians are varying with time.
(The most relevant work so far to our knowledge includes Ref. \cite{deClercq2015}
which uses basis splines to approximate a time-dependent Hamiltonian
of a qubit, and Ref. \cite{Tsang2011} which studies the quantum Cram\'er-Rao
bound for a time-varying signal, etc.) Nevertheless, in reality, many
factors that influence the systems are changing with time, e.g., periodic
driving fields or fluctuating external noise. In the state-of-the-art
field of quantum engineering, fast varying quantum controls are often
involved to improve operation fidelity and efficiency. Therefore,
the current knowledge about quantum metrology with static Hamiltonians
significantly limits application of quantum metrology in broader areas,
and the capability of treating time-dependent Hamiltonians is intrinsically
necessary for allowing the applicability of quantum metrology in more
complex situations.

In this article, we study quantum metrology with time-dependent Hamiltonians
to bridge this gap. We obtain the maximum quantum Fisher information
for parameters in time-dependent Hamiltonians in general, and show
that it is attainable only with proper control on the Hamiltonians
generally. The optimal Hamiltonian control and the measurement scheme
to achieve the maximum Fisher information are derived. Based on the
general results obtained, we surprisingly find that some fundamental
limits in quantum metrology with time-independent Hamiltonians can
be broken with time-dependent Hamiltonians. In a minimal example of
a qubit in a rotating magnetic field, we show that the time-scaling
of Fisher information for the rotation frequency of the field can
reach $T^{4}$ in the presence of the optimal Hamiltonian control,
significantly exceeding the traditional limit $T^{2}$ with time-independent
Hamiltonians. This suggests substantial differences between quantum
metrology with time-varying Hamiltonians and with static Hamiltonians.
Finally, we consider level crossings in the derivatives of Hamiltonians
with respect to the estimated parameters, and show that additional
Hamiltonian control is generally necessary to maximize the Fisher
information in that case.

\section*{Results}

\textbf{Quantum parameter estimation.} Parameter estimation is an
important task in vast areas of sciences, which is to extract the
parameter of interest from a distribution of data. The core goal of
parameter estimation is to increase the estimation precision. The
estimation precision is determined by how well the parameter can be
distinguished from a value in the vicinity, which can usually be characterized
by the statistical distance between the distributions with neighboring
parameters \cite{Wootters1981}. The well-known Cram\'er-Rao bound
\cite{Cramer1946} shows the universal limit of precision for arbitrary
estimation strategies, which indicates that for a parameter $g$ in
a probability distribution $\pkg$ of some random variable $X$, the
mean squared deviation $\langle\delta^{2}\eg\rangle\equiv{\rm E}\Big[\Big(\frac{\eg}{|\pg{\rm E}[\eg]|}-g\Big)^{2}\Big]$
is bounded by
\begin{equation}
\langle\delta^{2}\eg\rangle\geq\frac{1}{\nu\ig}+\langle\delta\eg\rangle^{2},
\end{equation}
where $\nu$ is the amount of data, $\ig$ is the Fisher information
\cite{Fisher1925},
\begin{equation}
\ig=\int\pkg(\pg\ln\pkg)^{2}\d X,\label{eq:classical fisher}
\end{equation}
and $\langle\delta\eg\rangle$ is the mean systematic error. For an
unbiased estimation strategy, $\langle\delta\eg\rangle=0$. The Cram\'er-Rao
bound can generally be achieved with the maximum likelihood estimation
strategy when the number of trials is sufficiently large \cite{Fisher1925}.
In practice, however, due to the finiteness of resource, only a limited
number of trials are available usually. For such situations, the Cram\'er-Rao
bound may become loose, and new families of error measures have been
proposed to give tighter bounds, for example Ref. \cite{Tsang2012}.
In this paper, we pursue the ultimate precision limit of quantum metrology
with time-dependent Hamiltonians allowed by quantum mechanics, regardless
of any practical imperfections like the finiteness of resources or
external noise, so the Cram\'er-Rao bound is the proper measure for
the estimation precision.

In the quantum regime of parameter estimation, we are interested in
estimating parameters in quantum states. The essence of estimating
a parameter in a quantum state is distinguishing the quantum state
with the parameter of interest from that state with a slightly deviated
parameter. When the quantum state is measured, the parameter in that
state controls the probability distribution of the measurement results,
and the information about the parameter can be extracted from the
measurement results. As there are many different possible measurements
on the same quantum state, the Fisher information needs to be maximized
over all possible measurements so as to properly quantify the distinguishability
of the quantum state with the parameter of interest. It is shown by
\cite{Braunstein1994,Braunstein1996} that the maximum Fisher information
for a parameter $g$ in a quantum state $\ksg$ over all possible
generalized quantum measurements is
\begin{equation}
\fq=4\big(\avg{\pg[g]\sg}-|\ovl{\sg}{\pg\sg}|^{2}\big).\label{eq:quantum fisher}
\end{equation}
This is called quantum Fisher information, and is closely related
to the Bures distance $\d s^{2}=2(1-|\ovl{\sg}{\psi_{g+\d g}}|)$
\cite{Bures1969} through $\d s^{2}=\frac{1}{4}\fq\d g^{2}$ between
two adjacent states $\ksg$ and $\ket{\psi_{g+\d g}}$, which characterizes
the distinguishability between $\ksg$ and $\ket{\psi_{g+\d g}}$.

\begin{figure}
\includegraphics[scale=0.47]{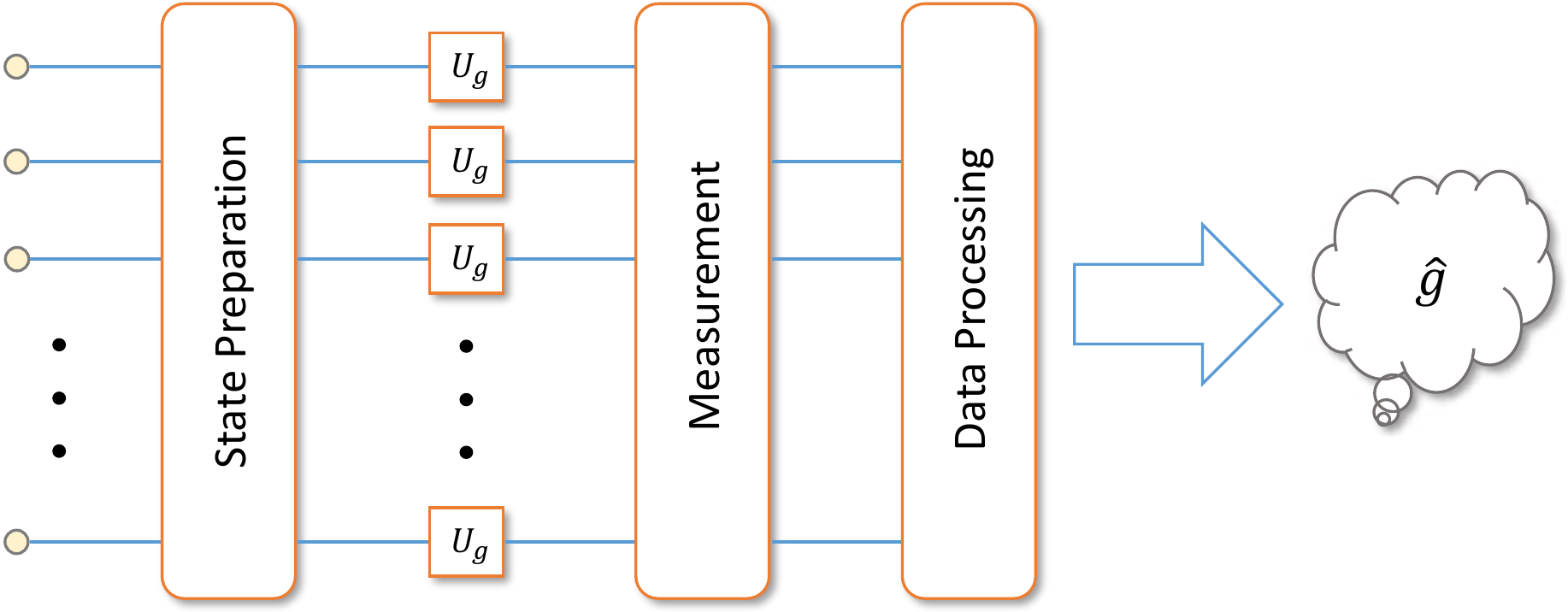}\caption{\label{fig:General-process-of}\textbf{General procedures of quantum
metrology.} Quantum metrology can generally be decomposed to four
steps: preparation of the initial states of the quantum systems, parameter-dependent
evolution ($\protect\ug$ in the figure) of the systems, measurements
on the final states of the systems, and post-processing of the measurement
data to extract the parameter. Each node at the left side of the figure
represents one quantum system (which can be very general and consist
of subsystems). Usually multiple systems are exploited to undergo
such a process, and they can be entangled at the preparation step
to increase the estimation precision beyond the standard quantum limit,
which is the advantage of quantum metrology.}
\end{figure}

In quantum metrology, the parameters to estimate are usually in Hamiltonians,
or more generally, in physical dynamics. The parameters are encoded
into quantum states by letting some quantum systems evolve under the
Hamiltonians or physical dynamics of interest. The states of the systems
acquire the information about the parameters from the evolution. The
parameters can then be learned from measurements on the final states
of the systems with appropriate processing of the measurement data.
A general process of quantum metrology is depicted in Fig. \ref{fig:General-process-of}.

A simple and widely-studied example of quantum metrology is to estimate
a multiplicative parameter in a Hamiltonian, say, to estimate $g$
in $\hgg=g\ho$ \cite{Giovannetti2006}, where $\ho$ is time-independent.
In this case, if a quantum systems undergoes the unitary evolution
$\ug=\exp(-\i g\ho T)$ for some time $T$, the quantum Fisher information
(\ref{eq:quantum fisher}) that determines the estimation precision
of $g$ is $\ig=4T^{2}\var[\ksg]{\ho}$, where $\var{\cdot}$ represents
variance and $\ksg$ is the final state of the system. A more general
case concerns a general parameter in a Hamiltonian \cite{Pang2014b}.
The quantum Fisher information for a general parameter $g$ in a Hamiltonian
$\hgg$ is $4\var[\ksg]{\hg}$, and $\h=\i(\pg\ug)\ug[\dagger]$ is
the local generator of the parametric translation of $\ug=\exp(-\i\hgg T)$
with respect to $g$ \cite{Braunstein1996}.

Compared to classical precision measurements, the advantage of quantum
metrology is that non-classical correlations can significantly enhance
measurement sensitivities. Various kinds of non-classical correlations
have been found useful for improving measurement precision, as reviewed
in the introduction. With $N$ properly correlated systems, the quantum
Fisher information can beat the standard quantum limit and attain
the Heisenberg scaling $N^{2}$ by appropriate metrological schemes
\cite{Giovannetti2004}.

\textbf{Time-dependent quantum metrology.} We now turn to the main
topic of this work, quantum metrology with time-dependent Hamiltonians.
Our goal is to find the maximum Fisher information for parameters
in time-dependent Hamiltonians.

The starting point of quantum metrology with a time-dependent Hamiltonian
is similar as with a time-independent Hamiltonian above. A system
is initialized in some state $\si$ and evolves under the time-dependent
Hamiltonian $\hgt$ with $g$ as the parameter to estimate, then after
an evolution for some time $T$, one measures the final state of the
system
\begin{equation}
\ksgt=\utg\si,\label{eq:evolved state}
\end{equation}
where $\utg$ is the unitary evolution under the Hamiltonian $\hgt$
for time $T$, and estimates $g$ from the measurement results, which
is just the standard recipe for a general quantum metrology. And the
quantum Fisher information of estimating $g$ from measuring $\ksgt$
is still determined by Eq. (\ref{eq:quantum fisher}), which can be
written as $\fq=4\var[\ksgt]{\hg}$, where $\hg=\i[\pg\utg[0]]\utgd$. 

Everything is similar as before so far, but we can immediately see
two major obstacles to deriving the maximum Fisher information. One
is that due to the complexity of evolution under a time-dependent
Hamiltonian, the unitary evolution $\utg$ is generally difficult
to obtain. The other is that even if we can find a solution to $\utg$,
it is hard to maximize the Fisher information, since $\hg$ can be
quite complex and the optimization is global involving the whole evolution
history of the system for time $T$. In order to derive the maximum
Fisher information for time-dependent Hamiltonians, we need to overcome
these obstacles.

For the purpose of convenience, we first reformulate the quantum Fisher
information as
\begin{equation}
\fq=4\var[\si]{\hg},\label{eq:20}
\end{equation}
which is dependent on the initial state $\si$ of the system now,
and $\hg$ becomes $\i\utgd\pg\utg[0]$, which is different from the
one in \cite{Braunstein1996,Pang2014b} and can no longer be interpreted
as the local generator of parametric translation of $\utg$ with respect
to $g$. But the maximum of the Fisher information $\fq$ is still
the squared gap between the maximum and minimum eigenvalues of $\hg$,
as in the case of static Hamiltonians \cite{Giovannetti2006}%
.%
{} Therefore, the key to determining the optimal estimation precision
for the parameter $g$ is finding $\hg$ and its maximum and minimum
eigenvalues.

Usually the evolution under a time-dependent Hamiltonian $\hgt$ is
represented by the time-ordered exponential of $\hgt$, but it is
complex and not convenient for our problem. Here we take an alternative
approach that breaks the unitary evolution $\utg$ into products of
small time intervals $\delt$ and takes the limit $\delt\rightarrow0$.
Interestingly, it turns out that with this approach, the maximum Fisher
information (and the optimal quantum control) can be obtained without
knowing the exact solution to $\utg$! We show in Supplementary Note
\ref{note1} that such an approach leads to
\begin{equation}
\hg=\int_{0}^{T}\utgd[0][][t]\phgt\utg[0][][t]\dt.
\end{equation}
Obviously, it still includes the unitary evolution $\utg[][][t]$
which is unknown. However, it has the advantage that it is an integral
over the time $t$, which makes it possible to decompose the global
optimization of the eigenvalues of $\hg$ into local optimizations
at each time point $t$. The idea is that, as is known, the maximum
eigenvalue of an Hermitian operator must be its largest expectation
value over all normalized states, so the maximum eigenvalue of $\hg$
is the maximum time integral of $\csi\utgd[0][][t]\phgt\utg[0][][t]\si$
from $0$ to $T$ over all $\si$. Considering $\utg$ is unitary,
$\utg[0][][t]\si$ is also a normalized state, so the upper bound
of $\csi\utgd[0][][t]\phgt\utg[0][][t]\si$ must be the maximum eigenvalue
of $\phgt$ at time $t$, which can be denoted as $\umax$. From this,
it can be immediately inferred that the maximum eigenvalue of $\hg$
is upper bounded by $\int_{0}^{T}\umax\dt$. Similarly, the minimum
eigenvalue of $\hg$ is lower bounded by $\int_{0}^{T}\umin\dt$.
With these two bounds for the maximum and minimum eigenvalues of $\hg$
respectively, we finally arrive at the upper bound of the quantum
Fisher information $\fq$,
\begin{equation}
\fq\leq\Big[\int_{0}^{T}(\umax-\umin)\dt\Big]^{2}.\label{eq:upper bound Fisher}
\end{equation}

It shows that the upper bound of the quantum Fisher information $\fq$
is determined by the integral of the gap between the maximum and minimum
eigenvalues of $\phgt$ from time $0$ to $T$. It can straightforwardly
recover the quantum Fisher information for a time-independent Hamiltonian
$\hgg$ by identifying $\umax$ at all times $t$ and identifying
$\umin$ at all times $t$, respectively. And when $\hgg=g\ho$, the
maximum Fisher information is just $T^{2}\Delta^{2}$, where $\Delta$
is the gap between the maximum and minimum eigenvalues of $\ho$,
the same as the result in \cite{Giovannetti2006}.

\textbf{Optimal Hamiltonian control.} A question that naturally arises
from the above result is whether the upper bound of quantum Fisher
information $\fq$ (\ref{eq:upper bound Fisher}) is achievable. From
the above derivation of the upper bound of $\fq$, it is obvious that
the upper bound cannot be saturated generally, unless there exists
initial states $\si$ and $\sl$ of the system such that $\utg[0][][t]\si$
and $\utg[0][][t]\sl$ are the instantaneous eigenstates of $\pg\hgt$
with the maximum and minimum eigenvalues, respectively, at any time
$t$. This imposes two conditions: (i) there exist $\si$ and $\sl$
which are the eigenstates of $\phgt[t]$ with the maximum and minimum
eigenvalues at the initial time $t=0$; (ii) $\utg[0][][t]\si$ and
$\utg[0][][t]\sl$ should remain as the eigenstates of $\phgt$ with
the maximum and minimum eigenvalues for all $t$ under the evolution
of $\hgt$. The first condition is easy to satisfy, but the second
one is difficult, since the time change of an instantaneous eigenstate
of $\phgt$ is generally different from the evolution under the Hamiltonian
$\hgt$ when $\hgt$ does not commute with $\phgt$ or $\hgt$ does
not commute between different time points. This condition is the main
obstacle to the saturation of the upper bound of Fisher information
(\ref{eq:upper bound Fisher}).

However, it inspires us to think that if we can add some control Hamiltonian,
which is independent of the parameter $g$, to the original Hamiltonian,
so that the state evolution under the total Hamiltonian is the same
as the time change of the instantaneous eigenstates of $\phgt$, then
a state starting from the eigenstate of $\phgt[0]$ with the maximum
or minimum eigenvalue will always stay in that eigenstate of $\phgt$
at any time $t$. And the upper bound of quantum Fisher information
$\fq$ can then be achieved by preparing the system in an equal superposition
of the eigenstates of $\phgt$ with the maximum and minimum eigenvalues
at the initial time $t=0$. So the key is finding such a control Hamiltonian.

A convenient way to realize the above target is to let each eigenstate
of $\phgt$ stay in the same eigenstate of $\phgt$ at all times $t$
when evolving under the total Hamiltonian. (Actually $\phgt$ should
be replaced by the derivative of total Hamiltonian now, but they are
the same because the control Hamiltonian must be independent of $g$.)
It implies that the $k$-th eigenstate $\klm[][k]$ of $\phgt$ should
satisfy the Schr\"odinger equation $\htot\klm[][k]=\i\pt\klm[][k]$,
where $\htot$ denotes the total Hamiltonian. Unlike the usual situations
where we know the Hamiltonian and want to find the solution to the
state, here we know the solution to the state, $\klm[][k]$, and need
to find the appropriate Hamiltonian $\htot$ that directs the evolution
instead. A simple solution to this equation is $\htot=\i\sum_{k}\klm[][k][\pt]\blm$.
(Note this solution is Hermitian because $\sum_{k}\klm[][k][\pt]\blm$
is skew-Hermitian.) Considering every eigenstate $\klm[][k]$ satisfies
the $U(1)$ symmetry, i.e., multiplying $\klm[][k]$ by an arbitrary
phase $\netk$ does not change that state, $\htot$ can be generalized
to include an additional term $\sum_{k}\thkd\klm[][k]\blm$. $\thkd$
can be replaced by arbitrary real functions $\fkt$, and $\thk=\int_{0}^{t}\fkt[\tp]\dtp$.
Thus, the optimal control Hamiltonian $\hc$ finally turns out to
be
\begin{equation}
\begin{aligned}\hc= & \sum_{k}\fkt\klm[][k]\blm-\hgt\\
 & +\i\sum_{k}\klm[][k][\pt]\blm.
\end{aligned}
\label{eq:control hamiltonian}
\end{equation}
It will be seen in the examples below that proper choices of the functions
$\fkt$ can significantly simplify the control Hamiltonian $\hc$
in some cases.

\begin{figure}
\includegraphics[scale=0.63]{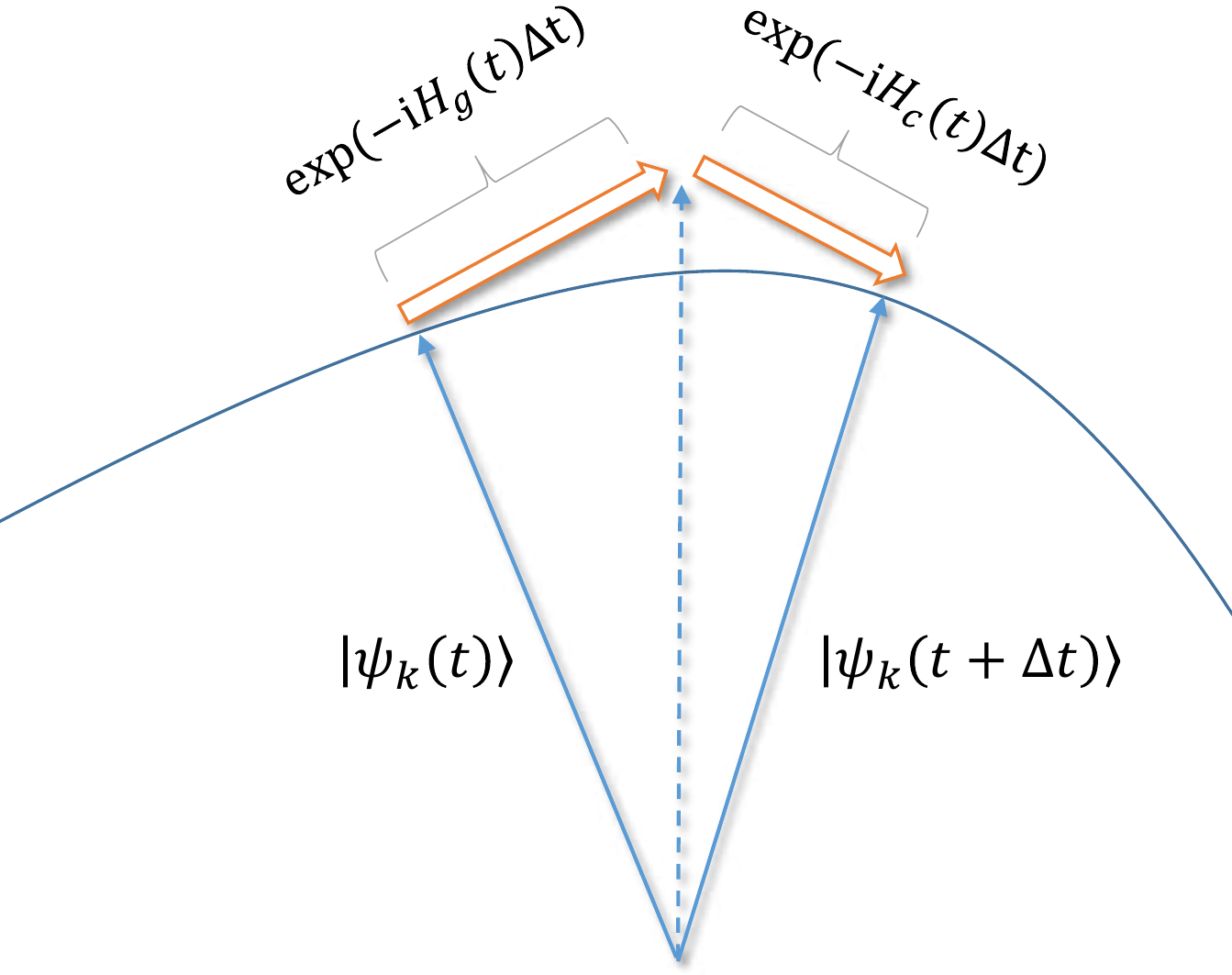}\caption{\label{fig:Optimal-Hamiltonian-control}\textbf{Optimal Hamiltonian
control scheme.} To achieve the maximum Fisher information, the optimal
control Hamiltonian needs to keep the eigenstates of $\protect\phgt$
evolving along the tracks of the eigenstates of $\protect\phgt$ under
the total Hamiltonian. The evolution under the total Hamiltonian $\protect\hgt+\protect\hc$
for a short time $\protect\delt$ can be approximated as $\exp(-\protect\i\protect\hc\protect\delt)\exp(-\protect\i\protect\hgt\protect\delt)$.
When $\exp(-\protect\i\protect\hgt\protect\delt)$ is applied on an
eigenstate $\protect\klm[][k]$ of $\protect\phgt$, the resulted
state (represented by the dashed ray in the figure) is not necessarily
still the instantaneous eigenstate $\protect\klm[t+\protect\delt][k]$
at time $t+\protect\delt$, and the role of the control Hamiltonian
$\protect\hc$ is to pull the state back to the instantaneous eigenstate
$\protect\klm[t+\protect\delt][k]$ at time $t+\protect\delt$. In
this way, with the assistance of the control Hamiltonian $\protect\hc$,
each eigenstate $\protect\klm[][k]$ of $\protect\phgt$ will always
evolve along the track of that eigenstate at any time $t$.}
\end{figure}

The role of this control Hamiltonian is to steer the eigenstates of
$\phgt$ evolving along the ``tracks'' of the eigenstates of $\phgt$
under the total Hamiltonian, which is the path to gain the most information
about $g$, instead of being deviated off the ``tracks'' by the
original Hamiltonian $\hgt$. This is critical to the saturation of
the upper bound of Fisher information. A schematic sketch for the
role of the optimal control Hamiltonian $\hc$ is plotted in Fig.
\ref{fig:Optimal-Hamiltonian-control}. It is worth mentioning that
similar ideas have been pursued in other works \cite{Garanin2002,Berry2009,Ruschhaupt2012,Barnes2013}
to steer the states of quantum systems along certain paths, such as
the instantaneous eigenstates of Hamiltonians, with proper control
fields.

It can be straightforwardly verified that with the above control Hamiltonian,
the eigenstates $\klm[0][k]$ of $\phgt$ at $t=0$ are the eigenstates
of $\hg[][t]$ for any time $t$, and the corresponding eigenvalues
are $\int_{0}^{T}\uk\dt$, where $\uk$ is the $k$-th eigenvalue
of $\phgt$ at time $t$. Therefore, $\hc$ indeed gives the demanded
control on the Hamiltonian to reach the upper and lower bounds of
the eigenvalues of $\hg$, and the upper bound of the quantum Fisher
information $\fq$ (\ref{eq:upper bound Fisher}) can then be achieved
by simply preparing the system in an equal superposition of the eigenstates
of $\phgt$ with the maximum and minimum eigenvalues at the initial
time $t=0$ and making proper measurements on the system after an
evolution of time $T$. The optimal measurement that gains the maximum
Fisher information is generally a projective measurement along the
basis $\kpm=\frac{1}{\sqrt{2}}(\e^{-\i\thk[T][\max]}\klmax[T]\pm\e^{-\i\thk[T][\min]}\klmin[T])$,
where $\klmax[T]$ and $\klmin[T]$ are the eigenstates of $\phgt$
with the maximum and minimum eigenvalues at time $t=T$, and $\thk[T][\max]$
and $\thk[T][\min]$ are the additional phases of $\klmax[T]$ and
$\klmin[T]$ depending on the choice of $\fkt$ in the optimal control
Hamiltonian (\ref{eq:control hamiltonian}). The details of the measurement
scheme are discussed in Supplementary Note \ref{note2}.

It is worth noting that the optimal control Hamiltonian (\ref{eq:control hamiltonian})
involves the estimated parameter $g$. However, $g$ is unknown, so
it should be replaced with a known estimate of $g$, say $\gc$, in
practice, and the control Hamiltonian becomes
\begin{equation}
\begin{aligned}\hc= & \sum_{k}\fkt|\widetilde{\psi_{k}}(t)\rangle\langle\widetilde{\psi_{k}}(t)|-\hgt[][\gc]\\
 & +\i\sum_{k}|\pt\widetilde{\psi_{k}}(t)\rangle\langle\widetilde{\psi_{k}}(t)|.
\end{aligned}
\label{eq:hcgc}
\end{equation}
where $g$ has been replaced by $\gc$ and $|\widetilde{\psi_{k}}(t)\rangle$
denotes the $k$-th eigenstate of $\phgt$ with $g=\gc$. 

The estimate $\gc$ can be first obtained by some estimation scheme
without the Hamiltonian control, then applied in the control Hamiltonian
to have a more precise estimate of $g$. The new estimate of $g$
can be fed back to the control Hamiltonian to further update the estimate
of $g$. Thus, the above Hamiltonian control scheme is essentially
adaptive, requiring feedback from each round of estimation to refine
the control Hamiltonian and optimize the estimation precision.

We stress that the control Hamiltonian (\ref{eq:hcgc}) is independent
of the parameter $g$, although the optimal control Hamiltonian (\ref{eq:control hamiltonian})
involves $g$, otherwise the control Hamiltonian would carry additional
information about $g$, which is not physical. From a quantum state
discrimination point of view, the estimation of $g$ is essentially
to distinguish between $\mt\exp[-\i\int_{0}^{T}\hgt\dt]\si$ and $\mt\exp[-\i\int_{0}^{T}\hgt[][g+\dg]\dt]\si$,
where $\si$ is the initial state of the system. When a control Hamiltonian
$\hc[t,\gc]$ is applied (where $\gc$ is explicitly denoted), the
two states become $\mt\exp[-\i\int_{0}^{T}(\hgt+\hc[t,\gc])\dt]\si$
and $\mt\exp[-\i\int_{0}^{T}(\hgt[][g+\dg]+\hc[t,\gc])\dt]\si$. One
can see that when $g$ has a virtual shift $\dg$ in the original
Hamiltonian, $\gc$ is unchanged in the control Hamiltonian. The parameter
$\gc$ in the control Hamiltonian is always a constant (even when
it is equal to the real value of $g$), while the parameter $g$ in
the original Hamiltonian is a variable. This is how the control Hamiltonian
is independent of $g$. The appearance of $g$ in the optimal control
Hamiltonian (\ref{eq:control hamiltonian}) just indicates what $\gc$
maximizes the Fisher information, and it turns out to be the real
value of $g$.

As a simple verification of the above results, we show how the current
results can recover the known ones in quantum metrology with time-independent
Hamiltonians. Consider estimating a multiplicative parameter $g$
in a time-independent Hamiltonian $\hgg=g\ho$, which is the simplest
case that has been widely studied. In this case, $\phgt=\ho$ and
$\klm[][k][\pt]=0$. To obtain a simple control Hamiltonian, we can
choose $\fkt$ to be the $k$-th eigenvalue $E_{k}$ of $\ho$, i.e.,
multiply $\klm[][k]$ with a phase $\e^{-\i E_{k}t}$, in the optimal
control Hamiltonian $\hc$ (\ref{eq:control hamiltonian}); then $\hc=0$.
This implies no Hamiltonian control is necessary for this case, in
accordance with the result in \cite{Giovannetti2006}.

A more general case is that the Hamiltonian is still independent of
time but the parameter to estimate is not necessarily multiplicative.
This has attracted a lot of attention recently \cite{Pang2014b,Yuan2015a,Jing2015,Liu2015,Liu2016,Yuan2016,Baumgratz2016}.
The Hamiltonian in this case can be represented as $\hgt=\hgg$ in
general. Since the Hamiltonian is still time independent, we have
$\klm[][k][\pt]=0$. So, the optimal control Hamiltonian is $\hc=\sum_{k}\fkt\klm[][k]\blm-\hgg$.
But in this case, $\klm[][k]$ are not necessarily the eigenstates
of $\hgg$, and $\sum_{k}\fkt\klm[][k]\blm$ cannot cancel $\hgg$
generally. To simplify $\hc$, we can simply choose $\fkt=0$, then
$\hc=-\hgg$. It implies a reverse of the original Hamiltonian can
lead to the maximum Fisher information in this case. This recovers
the result in \cite{Yuan2015a}, which showed that the optimal control
to maximize the quantum Fisher information for this case is just to
apply a reverse of the original unitary evolution at each time point.
Of course, this is not the unique solution to $\hc$, and a large
family of solutions exist corresponding to different choices of $\fkt$,
all leading to the maximum Fisher information.

\textbf{Estimation of field amplitude.} To exemplify the features
of quantum metrology with time-dependent Hamiltonians and the power
of the above Hamiltonian control scheme, we consider a simple physical
example below. This example will show some important characteristics
of time-dependent quantum metrology and how the optimized Hamiltonian
control can dramatically boost the estimation precision.

Let us consider a qubit in a uniformly rotating magnetic field, $\bt=B(\cos\omega t\ex+\sin\omega t\ez)$,
where $\ex$ and $\ez$ are the unit vectors in the $\hat{x}$ and
$\hat{z}$ directions, respectively, and we want to estimate the amplitude
$B$ or the rotation frequency $\omega$ of the field. To acquire
the information about the magnetic field, we let the qubit evolve
in the field for some time $T$, then measure the final state of the
qubit to learn $B$ or $\omega$. The interaction Hamiltonian $-\bt\cdot\sig$
between the qubit and the field is
\begin{equation}
\ham=-B(\cos\omega t\sx+\sin\omega t\sz),\label{eq:qubit hamiltonian}
\end{equation}
where we assumed the magnetic moment of the qubit to be $1$.

We first consider estimating the amplitude $B$ of the magnetic field.
It is easy to verify that the derivative of $\ham$ with respect to
$B$ has eigenvalues $\pm1$ for any $t$, therefore, the maximum
quantum Fisher information (\ref{eq:upper bound Fisher}) of estimating
$B$ at time $T$ is
\begin{equation}
\fq[B]=4T^{2}.\label{eq:18}
\end{equation}
As shown previously, it requires some control on the Hamiltonian to
reach this maximum quantum Fisher information. It can be straightforwardly
obtained that the eigenstates of $\pb\ham$ are $\klm[][+]=\cos\frac{\omega t}{2}\kp+\sin\frac{\omega t}{2}\km$
and $\klm[][-]=\sin\frac{\omega t}{2}\kp-\cos\frac{\omega t}{2}\km$,
where $\kpm=\frac{1}{\sqrt{2}}(\ko\pm\kl)$, corresponding to oscillations
in the $Z-X$ plane. Since $\pb\ham=B^{-1}\ham$, we can choose the
first term in Eq. (\ref{eq:control hamiltonian}) to cancel $\ham$.
Then, the optimal control Hamiltonian $\hc$ (\ref{eq:control hamiltonian})
is
\begin{equation}
\hc=-\frac{\omega}{2}\sy.\label{eq:b-ctrl}
\end{equation}

What about if we do not apply the control Hamiltonian $\hc$? We obtain
the evolution of the qubit and the quantum Fisher information for
the amplitude $B$ without any Hamiltonian control in Supplementary
Note \ref{note3}. The quantum Fisher information for this case is
\begin{equation}
\fq[B,0]=\frac{16B^{2}T^{2}}{4B^{2}+\omega^{2}}+8\omega^{2}\frac{1-\cos\left(T\sqrt{4B^{2}+\omega^{2}}\right)}{\left(4B^{2}+\omega^{2}\right)^{2}}.\label{eq:17}
\end{equation}
It implies that when $T\gg1$,
\begin{equation}
\frac{\fq[B]}{\fq[B,0]}\approx1+\frac{\omega^{2}}{4B^{2}},\label{eq:16}
\end{equation}
indicating that the increase of Fisher information by the Hamiltonian
control is determined by the ratio between $\omega$ and $B$. 

It is interesting to note that if the field rotation frequency $\omega$
is small, the increase in Fisher information by the Hamiltonian control
would be small as well, as shown by Eq. (\ref{eq:16}). This is because
when $\omega\ll1$, the magnetic field is changing so slowly that
the evolution of the qubit state is approximately adiabatic, and an
eigenstate of $\pb\ham$ would always stay in that eigenstate considering
$\pb\ham$ commutes with $\ham$. Thus, the condition for optimizing
the Fisher information can be automatically satisfied, and the maximum
Fisher information is achieved as a result. This is also verified
by Eq. (\ref{eq:b-ctrl}) that when $\omega\ll1$, the optimal control
Hamiltonian is close to zero, which means almost no Hamiltonian control
is necessary for this case.

\begin{figure}
\includegraphics[scale=0.66]{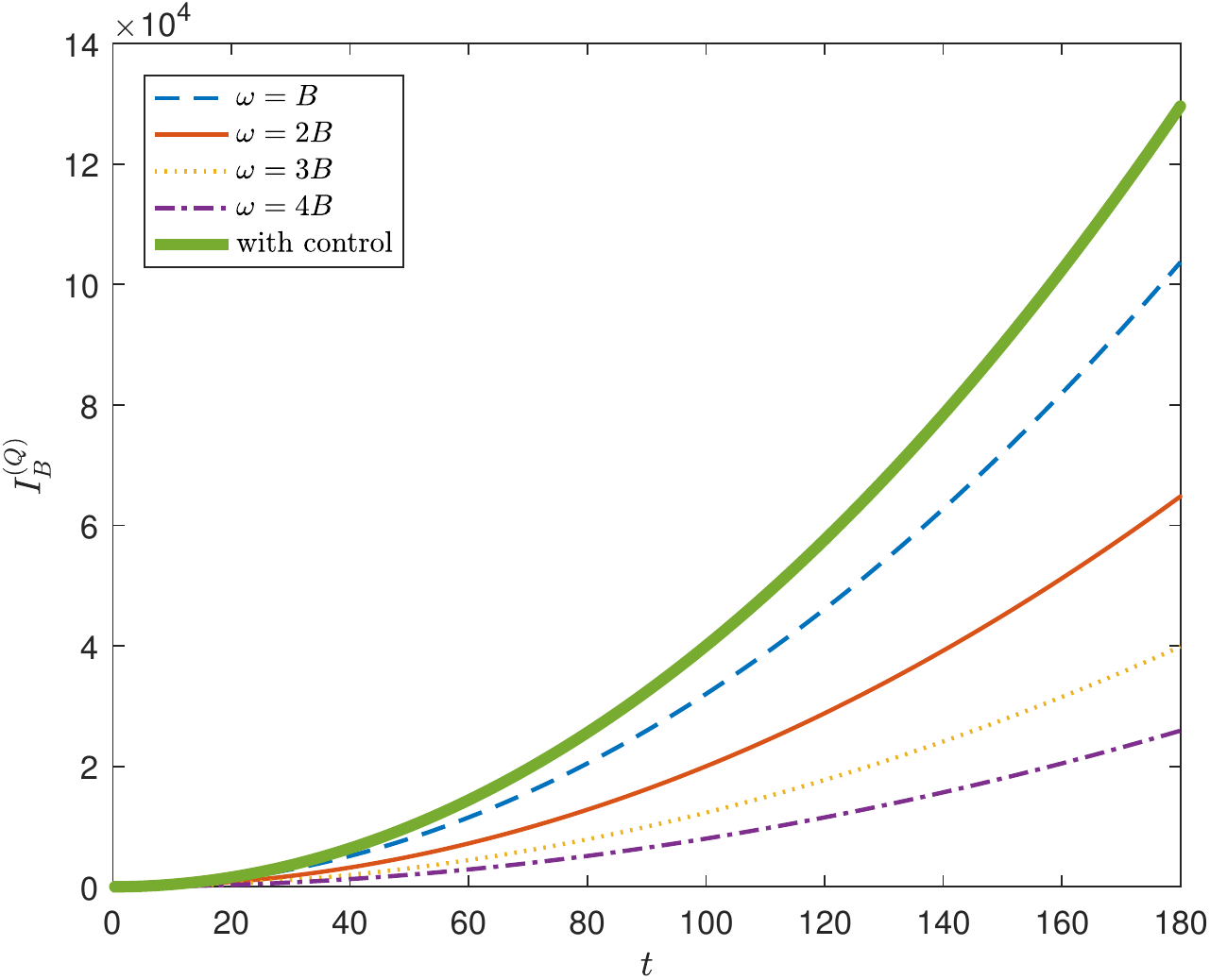}\caption{\label{fig:varying B}\textbf{Quantum Fisher information for the field
amplitude.} Quantum Fisher information $\protect\fq[B]$ for the amplitude
$B$ of the rotating magnetic field $\protect\bt$ versus the evolution
time $t$ is plotted for different choices of rotation frequency $\omega$
without the Hamiltonian control, and compared to that with the optimized
Hamiltonian control. The true value of $B$ in the figure is $1$.
It can be observed that when $\omega$ is large compared to the amplitude
of the magnetic field $B$, the Fisher information becomes small.
The Fisher information with the optimal Hamiltonian control is the
highest, whatever $\omega$ is, which verifies the advantage of Hamiltonian
control for this case.}
\end{figure}

The quantum Fisher information of $B$ is plotted for different rotation
frequencies $\omega$ without the control Hamiltonian and compared
to that with the optimal control Hamiltonian (\ref{eq:b-ctrl}) in
Fig. \ref{fig:varying B}.

\textbf{Estimation of field rotation frequency. }Now, we turn to the
estimation of the rotation frequency $\omega$ of the magnetic field.
Frequency measurement is important in many areas of physics, and has
been widely studied in different contexts, e.g., a single-spin spectrum
analyzer \cite{Kotler2013}. High precision phase estimation has been
realized in many experiments in recent years, for example, on a single
nuclear spin in diamond with a precision of order $T^{-0.85}$ by
Waldherr \emph{et al.} \cite{Waldherr2012}.

To study the estimation precision of the frequency $\omega$, note
$\po\ham$ is $tB(\sin\omega t\sx-\cos\omega t\sz)$. The eigenvalues
of $\po\ham$ are $\mu(t)=\pm tB$ then, so the maximum and minimum
eigenvalues of $\hg[\omega]$ are $\int_{0}^{T}\mu(t)\dt=\pm\frac{1}{2}BT^{2}$.
Therefore, the maximum Fisher information of estimating $\omega$
is
\begin{equation}
\fq[\omega]=B^{2}T^{4}.\label{eq:T^4}
\end{equation}
The eigenstates of $\po\ham$ are $\klm[][+]=\sin\frac{\omega t}{2}\ko+\cos\frac{\omega t}{2}\kl$
and $\klm[][-]=\cos\frac{\omega t}{2}\ko-\sin\frac{\omega t}{2}\kl$.
{} If we choose $\fkt[][k]=0$ for Eq. (\ref{eq:control hamiltonian}),
then the optimal control Hamiltonian is
\begin{equation}
\hc=B(\cos\omega t\sx+\sin\omega t\sz)-\frac{\omega}{2}\sy.\label{eq:ome-ctrl}
\end{equation}
The first term in $\hc$ (\ref{eq:ome-ctrl}) cancels the original
Hamiltonian $\ham$, so that the eigenstates of $\po\ham$ would not
be deviated by $\ham$, and the second term in $\hc$ guides the eigenstates
of $\po\ham$ along the tracks of those eigenstates during the whole
evolution under the total Hamiltonian with the control $\hc$.

The above result of $\fq[\omega]$ has an important implication: it
is known that the time scaling of Fisher information for a parameter
of a time-independent Hamiltonian is at most $T^{2}$, even with some
control on the Hamiltonian, a fundamental limit in time-independent
quantum metrology \cite{Yuan2015a}; however, in this example, the
time scaling of Fisher information for the frequency $\omega$ reaches
$T^{4}$, an order $T^{2}$ higher than the time-independent limit!
This indicates that some fundamental limits in the time-independent
quantum metrology no longer hold when the Hamiltonian becomes varying
with time, and they can be dramatically violated in the presence of
appropriate quantum control on the system, showing a significant discrepancy
between the time-dependent and the time-independent quantum metrology.

An interesting question that naturally arises is if there is no control
Hamiltonian $\hc$, can the maximum Fisher information $\fq[\omega]$
still scale as $T^{4}$? In Supplementary Note \ref{note3}, we derive
the maximum Fisher information for the rotation frequency $\omega$
in the absence of Hamiltonian control by an exact computation of the
qubit evolution in the rotating magnetic field, and the result turns
out to be
\begin{equation}
\begin{aligned}\fq[\omega,0]= & \frac{4B^{2}T^{2}}{4B^{2}+\omega^{2}}-\frac{8B^{2}T\sin\left(T\sqrt{4B^{2}+\omega^{2}}\right)}{\left(4B^{2}+\omega^{2}\right)^{3/2}}\\
 & +\frac{8B^{2}\left(1-\cos\left(T\sqrt{4B^{2}+\omega^{2}}\right)\right)}{\left(4B^{2}+\omega^{2}\right)^{2}}.
\end{aligned}
\end{equation}
Therefore, without any Hamiltonian control on the system, the Fisher
information would still scale as $T^{2}$ as in time-independent quantum
metrology, which is substantially lower than the $T^{4}$ scaling
with the optimized Hamiltonian control. This exhibits the advantage
of Hamiltonian control in enhancing time-dependent quantum metrology.%

\begin{figure}
\includegraphics[scale=0.66]{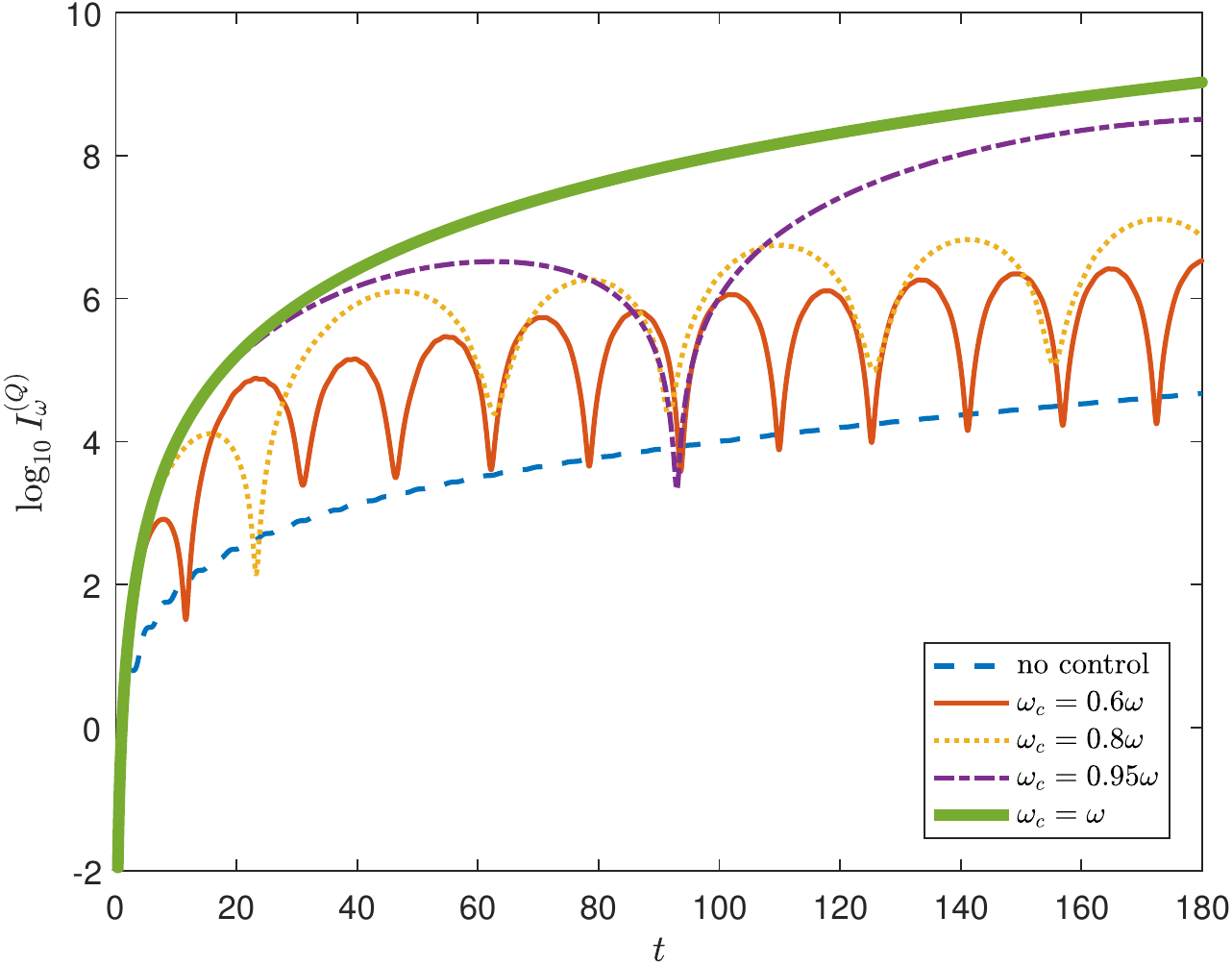}\caption{\label{fig:varying omega}\textbf{Quantum Fisher information for the
field frequency.} The logarithm (base 10) of the quantum Fisher information
$\protect\fq[\omega]$ for the rotation frequency $\omega$ of the
magnetic field $\protect\bt$ versus the evolution time $t$ is plotted
for different trial values $\protect\oc$ of the rotation frequency,
and compared to the Fisher information in the absence of the control
Hamiltonian $\protect\hc$. The real value of $B$ and the real value
of $\omega$ are both $1$. It can be observed that, even with some
sub-optimal choices of $\protect\oc$ which is not equal to the real
value of $\omega$, the scaling of the Fisher information can still
be much higher than that without any Hamiltonian control, and when
$\protect\oc$ approaches the real rotation frequency $\omega$ of
the magnetic field, higher Fisher information can be gained with the
assistance of Hamiltonian control. When $\protect\oc=\omega$, the
Fisher information reaches the maximum, which confirms the theoretical
results.}
\end{figure}

Fig. \ref{fig:varying omega} plots the Fisher information of $\omega$
in the presence of the control Hamiltonian with various $\oc$, and
compares it to that without the control Hamiltonian.

It should be noted that when the Hamiltonian is allowed to vary with
time, the time scaling of Fisher information may be raised in a trivial
way: the strength or the level gap of the Hamiltonian may itself increase
rapidly with time. For example, if the Hamiltonian is growing exponentially
with time (e.g., $\hgg=g\e^{t}\sz$), the Fisher information can have
an exponential time scaling. The nontriviality of the current result
lies in that the Hamiltonian (\ref{eq:qubit hamiltonian}) has a fixed
gap $2B$ between its highest and lowest levels, which does not scale
up with time, and thus the increase in Fisher information does not
result from any time growth of the Hamiltonian.

One may be wondering about the origin of the $T^{4}$ scaling. It
is not from the control Hamiltonian, since the control Hamiltonian
is independent of the estimated parameter, which is $\omega$ in this
example. The $T^{4}$ scaling originates from the dynamics of the
original Hamiltonian. Consider two original Hamiltonians with slightly
deviated parameters $\omega$ and $\omega+\do$. The discrepancy between
them is amplified by a time factor $t$ as they evolve, and correspondingly
the distance between the states evolving under these two Hamiltonians
is amplified by a time factor $t$ as well. Since the squared distance
between two states with neighboring parameters is approximately proportional
to the quantum Fisher information as manifested by the Bures metric
\cite{Bures1969}, the quantum Fisher information of $\omega$ can
therefore be increased by an order $T^{2}$ after an evolution of
time $T$. The control Hamiltonian helps keep the qubit on the optimal
route that gains the most Fisher information.

\textbf{Adaptive control for frequency estimation.} A notable point
in the above Hamiltonian control scheme for frequency estimation is
that the optimal control Hamiltonian $\hc$ (\ref{eq:ome-ctrl}) involves
the rotation frequency $\omega$. However, $\omega$ is the parameter
to estimate, so, in practice, we can only use an estimate of $\omega$,
say $\oc$, instead of the real value of $\omega$ in implementing
the control Hamiltonian (\ref{eq:ome-ctrl}), and the control Hamiltonian
would actually be
\begin{equation}
\hc=B(\cos\oc t\sx+\sin\oc t\sz)-\frac{\oc}{2}\sy.
\end{equation}
When the measurement runs for multiple rounds, the estimate $\oc$
will approach the real value of $\omega$, and the optimal Fisher
information (\ref{eq:T^4}) can be saturated by adaptively updating
the estimate of $\omega$ in the control Hamiltonian. This implies
that a feedback of the information about $\omega$ from each round
of measurement into the next round is necessary to implement the optimal
Hamiltonian control scheme and maximize the estimation precision for
$\omega$.

The details of the adaptive Hamiltonian control scheme are presented
in Supplementary Note \ref{note5}. Generally one needs to first obtain
an initial estimate of $\omega$ by some estimation scheme without
the Hamiltonian control, then apply it to the control Hamiltonian
and update it by estimation in the presence of the control Hamiltonian.
The updated estimate of $\omega$ can again be applied in the control
Hamiltonian to produce a better estimate of $\omega$, and so forth.

An important point shown in Supplementary Note \ref{note4} is that
with an estimate of $\omega$, $\oc$, which deviates from the exact
value of $\omega$ by $\do$, $\do=\oc-\omega$, the Fisher information
in the presence of the Hamiltonian control is approximately
\begin{equation}
\fq[\omega]=B^{2}T^{4}(1-\frac{1}{18}T^{2}\delta\omega^{2}).
\end{equation}
So, to approach the $T^{4}$ scaling of Fisher information for a given
evolution time $T$, the necessary precision $\do$ of the estimate
$\oc$ in the control Hamiltonian is only of the order $T^{-1}$,
so the feedback of a low precision estimate of $\omega$ in the Hamiltonian
control can lead to a high precision estimate of $\omega$. This lays
the foundation for the adaptive Hamiltonian control scheme. In particular,
it implies that the precision of the initial estimate of $\omega$
also just needs to be of the order $T^{-1}$, attainable in the absence
of Hamiltonian control, which is exactly what we need.

In fact, such an iterative feedback control scheme can approach the
$T^{4}$ scaling of Fisher information very efficiently. It is shown
in Supplementary Note \ref{note5} that the number of necessary rounds
of feedback control to realize the $T^{4}$ scaling for a large $T$
is only
\begin{equation}
n\sim\left\lceil \log_{2}\ln T\right\rceil ,
\end{equation}
a double logarithm of $T$, so very few rounds of feedback control
are necessary to approach the $T^{4}$ scaling.

It is also worth mentioning that there is a minimum precision requirement
of the initial estimation of $\omega$ without the Hamiltonian control
so that the Fisher information increases after each round of feedback
control:
\begin{equation}
I_{0}>\frac{1}{B^{2}(1-\frac{1}{18N})},
\end{equation}
where $N$ is the number of measurements in each round of feedback
control, otherwise the Fisher information would decrease as the feedback
control proceeds.

\textbf{Discussion.} The final problem we want to discuss about the
above optimal Hamiltonian control scheme for time-dependent quantum
metrology is the case that the eigenstate of $\phgt$ with the maximum
or minimum eigenvalue does not always stay in the same eigenstate
during the evolution. In deriving the optimal control Hamiltonian
(\ref{eq:control hamiltonian}), we let each eigenstate of $\phgt$
stay in the same eigenstate during the evolution for simplicity. This
implicitly assumes that the eigenstate of $\phgt$ with the maximum
or minimum eigenvalue also stays in the same eigenstate during the
evolution. However, if the highest or lowest level crosses other levels
of $\phgt$, the corresponding eigenstate will change from one eigenstate
of $\phgt[t]$ to another at the crossing.

In the presence of such a level crossing, the upper bound of the maximum
eigenvalue of $\hg$ or the lower bound of the minimum eigenvalue
of $\hg$ cannot be attained, and as a result the upper bound (\ref{eq:upper bound Fisher})
on the quantum Fisher information cannot be saturated. In particular,
if the highest and lowest levels of $\phgt$ cross each other, the
Fisher information will even drop after the crossing, because the
gap between the maximum and minimum eigenvalues of $\hg$ will shrink.
Thus, it is necessary to cancel or suppress the effect of level crossing
in $\phgt$ in order to maximize the Fisher information.

In order to keep the highest or lowest level of $\phgt$ still in
the the highest or lowest level after a crossing in $\phgt$, we need
to change the dynamics of the system near the crossing so that the
highest or lowest level of $\phgt$ before the level crossing transits
to the new one after the level crossing. %
{} We propose an additional Hamiltonian control scheme in the Methods
to realize such a transition.

\begin{figure}
\raggedright{}\includegraphics[scale=0.6]{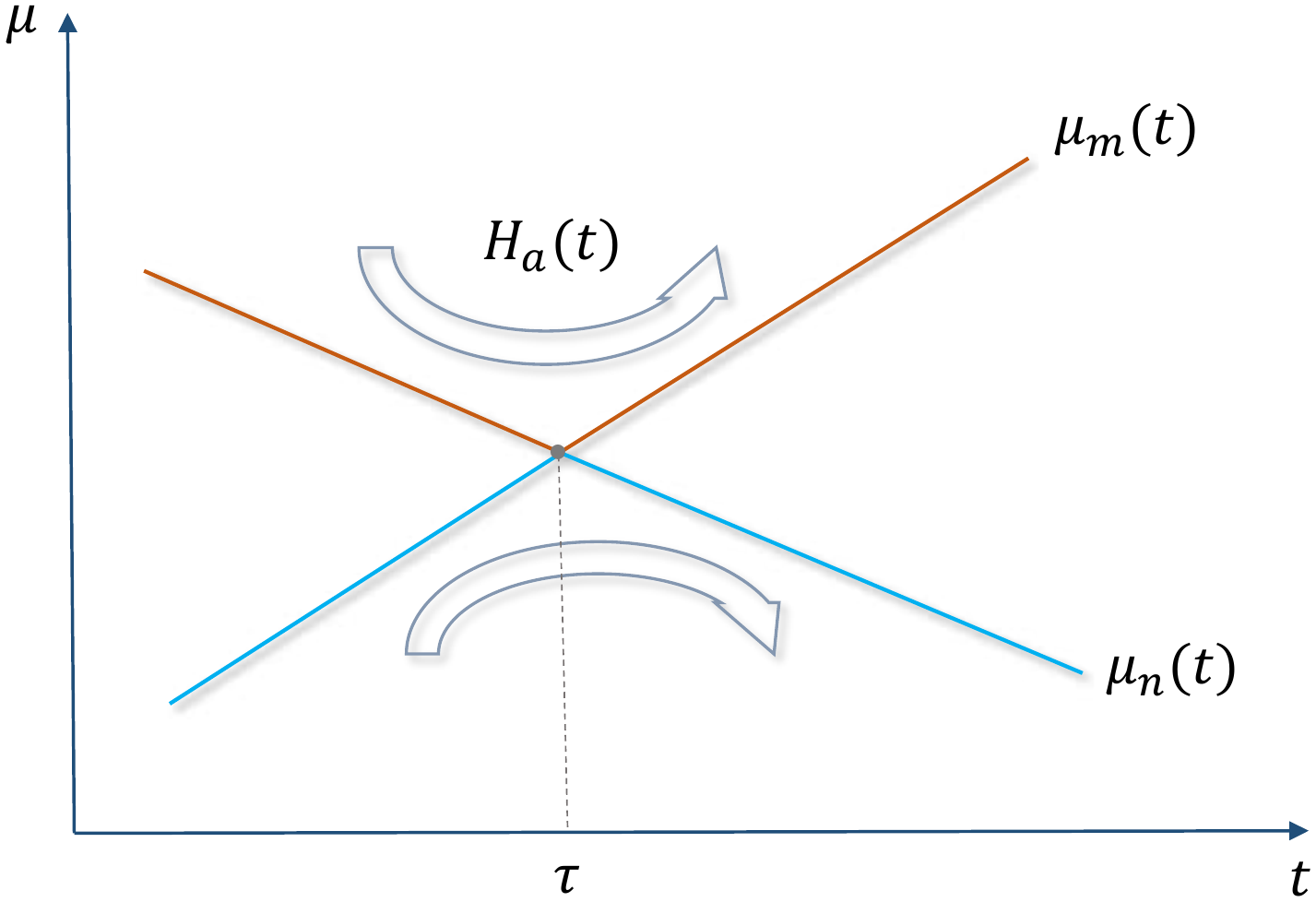}\caption{\label{fig:Additional-Hamiltonian-control}\textbf{Additional Hamiltonian
control scheme at level crossings of} $\boldsymbol{\protect\phgt}$\textbf{.}
Additional Hamiltonian control is necessary to eliminate the effect
of a crossing between the highest/lowest level and another level of
$\protect\phgt$. The role of the additional control Hamiltonian $\protect\ha$
is to transform the original instantaneous highest/lowest level to
the new instantaneous highest/lowest level of $\protect\phgt$. Suppose
the red curve in the figure is the highest level of $\protect\phgt$.
Before $\protect\tk$, $\protect\uk[][n]$ is the highest level of
$\protect\phgt$. At time $\protect\tk$, $\protect\uk[][n]$ crosses
the level $\protect\uk[][m]$, which becomes the highest level after
the crossing. The additional control Hamiltonian $\protect\ha$ is
to transit the highest level from $\protect\uk[][n]$ to $\protect\uk[][m]$
at time $\protect\tk$. The argument is similar if the blue curve
is the lowest level of $\protect\phgt$.}
\end{figure}

In Fig. \ref{fig:Additional-Hamiltonian-control}, the role of the
additional Hamiltonian control is plotted. When there are multiple
crossings between the highest/lowest level and other levels of $\phgt$
during the whole evolution process, there must be an additional Hamiltonian
control applied at each level crossing.

\section*{Methods}

\textbf{Additional quantum control at level crossings of $\boldsymbol{\phgt}$.}
Suppose a crossing occurs between the highest or lowest level and
another level of $\phgt$ at time $\tk$. $\uk[][n]$ is the highest
or lowest level of $\phgt$ before $\tk$ while $\uk[][m]$ becomes
the highest or lowest level after $\tk$, and $\klm[][n]$ and $\klm[][m]$
are the corresponding eigenstates. Intuitively, the following $\sx$-like
control Hamiltonian
\begin{equation}
\begin{aligned}\ha= & \hkt{}[\etkj[m][n]\klm[][n]\blm[][m]\\
 & +\etkj[n][m]\klm[][m]\blm[][n]],
\end{aligned}
\label{eq:additional Hamiltonian}
\end{equation}
should rotate $\klm[][n]$ to $\klm[][m]$, with $\hkt$ to be some
time-dependent control parameters and $\netk[m]$, $\netk[n]$ to
be the additional phases of $\klm[][m]$, $\klm[][n]$ determined
by the choices of $\fkt[][m]$, $\fkt[][n]$ in the optimal control
Hamiltonian (\ref{eq:control hamiltonian}). In order not to affect
the additional Hamiltonian controls at other level crossings, $\ha$
must be completed within a sufficiently short time $\sdt$. As shown
in Supplementary Note \ref{note6}, the control parameter $\hkt$
must satisfy
\begin{equation}
\int_{\tk-\frac{1}{2}\sdt}^{\tk+\frac{1}{2}\sdt}\hkt\dt=(l+\frac{1}{2})\pi,\label{eq:ht-cond}
\end{equation}
where $l$ is an arbitrary integer, so that the system can be exactly
transferred to the new eigenstate $\klm[][m]$ from $\klm[][n]$ by
the additional control Hamiltonian.

An intuitive idea why the above additional control Hamiltonian $\ha$
can drive $\klm[][n]$ to $\klm[][m]$ can be understood as follows.
Note that the total Hamiltonian is the sum of $\hgt$, $\hc$ and
the additional control Hamiltonian $\ha$ now. According to the time-dependent
generalization of the Suzuki-Trotter product formula \cite{Poulin2011},
if we break the time interval $\tk-\frac{1}{2}\sdt\leq t\leq\tk+\frac{1}{2}\sdt$
into many small pieces at properly sampled time points $t_{1},\,t_{2},\,\cdots,\,t_{n}$,
the total evolution of the system from $\tk-\frac{1}{2}\sdt$ to $\tk+\frac{1}{2}\sdt$
can be approximated as the time-ordered product of $\utg[t_{j}][][t_{j+1}]\approx\exp[-\i(\hgt[t_{j}]+\hc[t_{j}])\delt_{j}]\exp[-\i\ha[t_{j}]\delt_{j}]$,
where $\delt_{j}=t_{j+1}-t_{j}$, implying that at each short time
piece $\delt_{j}$, the state $\netk[n][t_{j}]\klm[t_{j}][n]$ is
slightly shifted to $\netk[m][t_{j}]\klm[t_{j}][m]$ by $\ha[t_{j}]$,
following which $\netk[n][t_{j}]\klm[t_{j}][n]$ is shifted to $\netk[n][t_{j+1}]\klm[t_{j+1}][n]$
and $\netk[m][t_{j}]\klm[t_{j}][m]$ is shifted to $\netk[m][t_{j+1}]\klm[t_{j+1}][m]$
by $\hgt[t_{j}]+\hc[t_{j}]$. Thus, the total effect of the additional
control Hamiltonian $\ha$, along with the original Hamiltonian $\hgt$
and the control Hamiltonian $\hc$, is continuously driving the system
from $\netk[n]\klm[][n]$ to $\netk[m]\klm[][m]$, where $\klm[][n]$
and $\klm[][m]$ are also changing at the same time. 

A rigorous analysis for the additional Hamiltonian $\ha$ is given
in Supplementary Note \ref{note6}. It turns out that in a rotating
frame where all $\netk\klm[][k]$ are static, the total Hamiltonian
is transformed to $\hp=\hkt\smn$, where $\smn$ is a $\sx$-like
transition operator between two static basis states $\kk[n]$ and
$\kk[m]$ in the new frame which correspond to $\netk[n]\klm[][n]$
and $\netk[m]\klm[][m]$ in the original frame. This indicates that
in the presence of the additional control Hamiltonian $\ha$, $\klm[][n]$
can be transited to $\klm[][m]$ continuously around the level crossing
between $\uk[][n]$ and $\uk[][m]$.

It should be noted that an additional phase $(-1)^{l+1}\i$ will be
introduced to the eigenstates $\klm[][m]$ and $\klm[][n]$ of $\phgt$
by the additional Hamiltonian control. This may change the relative
phase of the system when it is in a superposed state involving $\klm[][m]$
or $\klm[][n]$ and needs to be taken into account in that case. The
detail about the additional phase is given in Supplementary Note \ref{note6}.

\section*{Data availability}

The code and data used in this work are available upon request to
the corresponding author.

\section*{Acknowledgments}

The authors acknowledge the support from the US Army Research Office
under Grant No. W911NF-15-1-0496, W911NF-13-1-0402 and the support
from the National Science Foundation under Grant No. DMR-1506081.

\section*{Author contributions}

S.P. initiated this work, and carried out the main calculations. A.N.J.
participated in scientific discussions, and assisted with the calculations.
Both authors contributed to the writing of the manuscript.

\section*{Competing financial interests}

The authors declare no competing financial interests.

\bibliographystyle{naturemag}
\bibliography{optimal_control}

\newpage \setcounter{equation}{0} \newcounter{suppnote} \renewcommand{\theequation}{S\arabic{equation}} \onecolumngrid \setcounter{enumiv}{0}

\refstepcounter{suppnote}

\section*{Supplementary Note 1. Derivation of $\boldsymbol{\protect\hg}$\label{note1}}

In this Supplementary Note, we derive $\hg$ for a time-dependent
Hamiltonian $\hgt$, defined as
\begin{equation}
\hg=\i\utgd\pg\utg,\label{eq:hgt}
\end{equation}
where $\utg$ is the unitary evolution under the time-dependent Hamiltonian
$\hgt$, and it determines the quantum Fisher information of estimating
$g$ in the following way,
\begin{equation}
\fq=4\csi\var{\hg}\si,\label{eq:fish}
\end{equation}
where $\si$ is the initial state of the system. The maximum quantum
Fisher information is the square of the gap between the maximum and
minimum eigenvalues of $\hg$.

In order to obtain $\hg$, we break the unitary evolution $\utg$
for a time duration $T$ into small time intervals $\delt$,
\begin{equation}
\utg=\utg[T-\delt]\utg[T-2\delt][][T-\delt]\cdots\utg[\delt][][2\delt]\utg[][][\delt].
\end{equation}
Then,
\begin{equation}
\begin{aligned}\pg\utg= & \sum_{k=0}^{T/\delt-1}\Big\{\utg[T-\delt]\cdots\utg[(k+1)\delt][][(k+2)\delt]\\
 & [\pg\utg[k\delt][][(k+1)\delt]]\utg[(k-1)\delt][][k\delt]\cdots\utg[0][][\delt]\Big\}.
\end{aligned}
\end{equation}

When the time interval $\delt$ is sufficiently small, the Hamiltonian
$\hgt$ is approximately time-independent during each time interval
$k\delt\leq t\leq(k+1)\delt$, i.e.,
\begin{equation}
\utg[k\delt][][(k+1)\delt]\approx\exp(-\i\hgt[k\delt]\delt),
\end{equation}
and by expanding $\exp(-\i\hgt[k\delt]\delt)$ to the first order
of $\delt$, we have
\begin{equation}
\exp(-\i\hgt[k\delt]\delt)=I-\i\delt\hgt[k\delt]+O(\delt^{2}),
\end{equation}
so,
\begin{equation}
\pg\utg[k\delt][][(k+1)\delt]\approx-\i\delt\pg\hgt[k\delt]+O(\delt^{2}).
\end{equation}
In the limit $\delt\rightarrow0$, $\pg\utg$ can be written in the
following integral form,
\begin{equation}
\pg\utg=-\i\int_{0}^{T}\utg[t]\phgt\utg[][][t]\dt,\label{eq:dgut}
\end{equation}
which is the exact solution to $\pg\utg$.

When $\utgd$ is multiplied to $\pg\utg$ from the left, since
\begin{equation}
\utgd\utg[t]=\utgd[][][t],
\end{equation}
$\hg$ in Eq. (\ref{eq:hgt}) turns out to be
\begin{equation}
\hg=\int_{0}^{T}\utgd[0][][t]\phgt\utg[0][][t]\dt.\label{eq:2}
\end{equation}
This gives the integral form of $\hg$ in the main text for a time-dependent
Hamiltonian $\hgt$ at time $t=T$. When $\hgt$ is independent of
time, Eq. (\ref{eq:2}) degenerates to the relevant formula in \cite{Pang2014b}.

\refstepcounter{suppnote}

\section*{Supplementary Note 2. Measurement scheme for estimation with optimal
Hamiltonian control\label{note2}}

In this Supplementary Note, we derive the estimator and the measurement
scheme that can gain the upper bound of Fisher information given by
Eq. (7) in the main manuscript with the assistance of the optimal
Hamiltonian control.

Suppose the parameter that we want to estimate is $g$. According
to Eq. (8) of the main manuscript, the optimal control Hamiltonian
is
\begin{equation}
\hc=\sum_{k}\fkt\klm[][k]\blm-\hgt+\i\sum_{k}\klm[][k][\pt]\blm,\label{eq:ctrl-ham}
\end{equation}
where $\hgt$ is the original Hamiltonian with the parameter $g$,
$\klm[][k]$ are the eigenstates of $\phgt$ and $\fkt$ are arbitrary
real functions. Since we do not know the exact value of $g$, the
parameter $g$ in the optimal control Hamiltonian (\ref{eq:ctrl-ham})
should be replaced by some known estimate of $g$, $\gc$, in practice,
and the control Hamiltonian is actually
\begin{equation}
\hc=\sum_{k}\fkt\ktlm\btlm-\hgt[][\gc]+\i\sum_{k}\ktlm[\pt]\btlm,\label{eq:approx-ctrl-ham}
\end{equation}
where $\ktlm$ are also dependent on $\gc$ instead of $g$, i.e.,
$\ktlm$ are eigenstates of $\phgt|_{g=\gc}$.

It should be noted that the parameter $\gc$ in the control Hamiltonian
(\ref{eq:approx-ctrl-ham}) is always a constant, even when it is
equal to the real value of $g$, while the parameter $g$ in the original
Hamiltonian $\hgt$ is a variable. This should be kept in mind in
computing the Fisher information for $g$. The appearance of $g$
in the optimal control Hamiltonian (\ref{eq:ctrl-ham}) just indicates
what value of $\gc$ maximizes the Fisher information, and it turns
out to be the real value of $g$.

When $\gc$ is close to $g$, the total Hamiltonian can be written
as
\begin{equation}
\begin{aligned}\htot= & \hgt+\hc\\
= & \sum_{k}\fkt\ktlm\btlm+\phgt|_{g=\gc}\dg+\i\sum_{k}\ktlm[\pt]\btlm,
\end{aligned}
\end{equation}
up to the first order of $\dg$, where $\dg=g-\gc$.

The evolution under $\htot$ can be decomposed as
\begin{equation}
\ut=\lim_{\delt\rightarrow0}\exp(-\i\htot[T]\delt)\exp(-\i\htot[T-\delt]\delt)\cdots\exp(-\i\htot[0]\delt),\label{eq:u0t}
\end{equation}
and at time $t$,
\begin{equation}
\exp(-\i\htot[t]\delt)\approx\exp(-\i\sum_{k}\ktlm[\pt]\btlm\delt)\exp(-\i\sum_{k}\fkt\ktlm\btlm\delt)\exp(-\i\phgt|_{g=\gc}\dg\delt).\label{eq:exphtott}
\end{equation}
Note that generally different orderings of the three terms at the
right side of Eq. (\ref{eq:exphtott}) will give different results
when $\delt$ is finite, but when $\delt\rightarrow0$, they will
give the same result. The ordering chosen in (\ref{eq:exphtott})
is for convenience of computation below.

If a system is initially in an eigenstate of $\phgt|_{g=\gc}$ at
$t=0$, say $\ktlm[][0]$, then according to Eq. (\ref{eq:u0t}) and
(\ref{eq:exphtott}), the state after an evolution of time $T$ is
\begin{equation}
\ut\ktlm[][0]=\exp\Big[-\i\Big(\thk[T]+\dg\int_{0}^{T}\uk\dt\Big)\Big]\ktlm[][T],
\end{equation}
where
\begin{equation}
\thk[T]=\int_{0}^{T}\fkt\dt.
\end{equation}

Now, suppose the maximum and minimum eigenvalues of $\phgt|_{g=\gc}$
at time $t$ are $\umax$ and $\umin$, and the corresponding eigenstates
are $\ktlmax$ and $\ktlmin$, respectively. To achieve the maximum
Fisher information given by Eq. (7) in the main manuscript, we can
prepare the system in an equal superposition of $\ktlmax[][0]$ and
$\ktlmin[][0]$ at the initial time $t=0$,
\begin{equation}
\pst[0]=\frac{1}{\sqrt{2}}(\ktlmax[][0]+\ktlmin[][0]).
\end{equation}
Then after evolving for time $T$, the state of the system is
\begin{equation}
\begin{aligned}\pst[T]= & \frac{1}{\sqrt{2}}\Big\{\exp\Big[-\i\Big(\thk[T][\max]+\dg\int_{0}^{T}\umax\dt\Big)\Big]\ktlmax[][T]\\
 & +\exp\Big[-\i\Big(\thk[T][\min]+\dg\int_{0}^{T}\umin\dt\Big)\Big]\ktlmin[][T]\Big\}.
\end{aligned}
\end{equation}
Note that $\thk[][\max/\min]$ denotes the value of $\thk$ associated
with the maximum or minimum eigenvalue of $\phgt|_{g=\gc}$, but not
the maximum or minimum over $\thk$.

Since $\pg=\pg[\dg]$, the quantum Fisher information of estimating
of $g$ by measuring $\pst[T]$ is
\begin{equation}
\begin{aligned}\fq= & 4\Big(\avg{\pg[\dg]\Psi(T)}-|\ovl{\Psi(T)}{\pg[\dg]\Psi(T)}|^{2}\Big)\\
= & \Big[\int_{0}^{T}\big(\umax-\umin\big)\dt\Big]^{2},
\end{aligned}
\end{equation}
which is exactly the upper bound of quantum Fisher information given
by Eq. (7) in the main manuscript.

To attain this quantum Fisher information, we can measure the following
observable
\begin{equation}
\mo=\kp\bp-\km\bmm,
\end{equation}
where
\begin{equation}
\kpm=\frac{1}{\sqrt{2}}(\e^{-\i\thk[T][\max]}\ktlmax[][T]\pm\e^{-\i\thk[T][\min]}\ktlmin[][T]),
\end{equation}
which are varying with time $T$. It is important to note that $\kpm$
are also dependent on $\gc$, instead of $g$, which makes it possible
to implement the measurement of $\mo$ without knowing the exact value
of $g$.

It can be obtained that the expectation value and variance of $\mo$
under the state $\pst$ are
\begin{equation}
\begin{aligned}\av{\mo}= & \cos\Big[\int_{0}^{T}(\umax-\umin)\dg\dt\Big],\\
\av{\Delta\mo^{2}}= & \sin^{2}\Big[\int_{0}^{T}(\umax-\umin)\dg\dt\Big].
\end{aligned}
\end{equation}
$\av{\mo}$ can be considered as an estimator of $g$ (with some local
unit difference characterized by $\pg[\dg]\av{\mo}$ and potential
systematic errors which can be eliminated by calibration) since it
is dependent on $g$. The parameter $g$ can be obtained from $\av{\mo}$,
and the variance of the estimate \cite{Braunstein1994,Giovannetti2006}
is
\begin{equation}
\delta g^{2}=\frac{\av{\Delta\mo^{2}}}{|\pg[\dg]\av{\mo}|^{2}}=\frac{1}{{\displaystyle \Big[\int_{0}^{T}(\umax-\umin)\dt\Big]^{2}}},\label{eq:dg}
\end{equation}
which exactly saturates the upper bound of Fisher information in Eq.
(7) of the main manuscript. If there are $N$ trials, the precision
would be $1/\sqrt{N}$ of $\delta g$ then.

\refstepcounter{suppnote}

\section*{Supplementary Note 3. Fisher information for $\boldsymbol{B}$ and
$\boldsymbol{\omega}$ in the absence of Hamiltonian control\label{note3}}

In this Supplementary Note, we derive the evolution of a qubit in
a rotating magnetic field and the optimal quantum Fisher information
for the amplitude $B$ and the rotation frequency $\omega$ of the
magnetic field in the absence of control Hamiltonian.

Suppose the rotating magnetic field is $\bt$,
\begin{equation}
\bt=B(\cos\omega t\ex+\sin\omega t\ez),\label{eq:bt}
\end{equation}
where the amplitude of the field $B$ is assumed to be constant for
simplicity. The interaction Hamiltonian between a qubit and the field
is
\begin{equation}
\ham=-\bt\cdot\sig=-B(\cos\omega t\sx+\sin\omega t\sz).\label{eq:rotham}
\end{equation}
When there is no control on the Hamiltonian, the evolution of the
qubit is determined by the Schr\"odinger equation
\begin{equation}
\i\pt\pst=\ham\pst.\label{eq:schrodinger eq}
\end{equation}

To derive the evolution of the qubit in this case, note that
\begin{equation}
\exp(\i\frac{\omega}{2}\sy t)\sx\exp(-\i\frac{\omega}{2}\sy t)=\cos\omega t\sx+\sin\omega t\sz.
\end{equation}
We know that if a state under a general Hamiltonian $\ham$ is transformed
by $\exp(\i\ho t)$, the effective Hamiltonian for the evolution of
the new state is
\begin{equation}
\hp=\exp(\i\ho t)\ham\exp(-\i\ho t)-\ho.
\end{equation}
Therefore, the rotating Hamiltonian (\ref{eq:rotham}) can be perceived
as the effective Hamiltonian of an original Hamiltonian $-B\sx+\frac{\omega}{2}\sy$
in a frame rotating as $\exp(\i\frac{\omega}{2}\sy t)$.

The evolution of a qubit under the Hamiltonian $-B\sx+\frac{\omega}{2}\sy$
is $\exp[\i(B\sx-\frac{\omega}{2}\sy)t]$, so the evolution under
the rotating Hamiltonian (\ref{eq:rotham}) is
\begin{equation}
\ut[][t]=\exp(\i\frac{\omega}{2}\sy t)\exp[\i(B\sx-\frac{\omega}{2}\sy)t].
\end{equation}

Having obtained the evolution of the qubit, we can compute the Fisher
information for $B$ and $\omega$.

To obtain the optimal quantum Fisher information for the field amplitude
$B$, we calculate
\begin{equation}
\pb\ham=-(\cos\omega t\sx+\sin\omega t\sz).
\end{equation}

The optimal quantum Fisher information is determined by the squared
gap between the maximum and the minimum eigenvalues of
\begin{equation}
\begin{aligned}\hg[B]= & \int_{0}^{T}\utd[][t]\pb\ham\ut[][t]\dt\\
= & -\left(\frac{4B^{2}T}{4B^{2}+\omega^{2}}+\frac{\omega^{2}\sin\left(T\sqrt{4B^{2}+\omega^{2}}\right)}{\left(4B^{2}+\omega^{2}\right)^{3/2}}\right)\sx+2B\omega\left(\frac{T}{4B^{2}+\omega^{2}}-\frac{\sin\left(T\sqrt{4B^{2}+\omega^{2}}\right)}{\left(4B^{2}+\omega^{2}\right)^{3/2}}\right)\sy\\
 & -\frac{\omega\left(1-\cos\left(T\sqrt{4B^{2}+\omega^{2}}\right)\right)}{4B^{2}+\omega^{2}}\sz.
\end{aligned}
\end{equation}
From Eq. (\ref{eq:fish}), it can be obtained that the optimal quantum
Fisher information for $B$ is
\begin{equation}
\fq[B]=\frac{16B^{2}T^{2}}{4B^{2}+\omega^{2}}+8\omega^{2}\frac{1-\cos\left(T\sqrt{4B^{2}+\omega^{2}}\right)}{\left(4B^{2}+\omega^{2}\right)^{2}}.
\end{equation}

Similarly, to obtain the optimal quantum Fisher information for the
field rotation frequency $\omega$, we calculate
\begin{equation}
\po\ham=tB(\sin\omega t\sx-\cos\omega t\sz).
\end{equation}
And the $h$ matrix for $\omega$ is
\begin{equation}
\begin{aligned}\hg[\omega]= & \int_{0}^{T}\utd[][t]\po\ham\ut[][t]\dt\\
= & B\left(\frac{\sin\left(T\sqrt{4B^{2}+\omega^{2}}\right)}{\left(4B^{2}+\omega^{2}\right)^{3/2}}-\frac{T\cos\left(T\sqrt{4B^{2}+\omega^{2}}\right)}{4B^{2}+\omega^{2}}\right)(\omega\sx+2B\sy)\\
 & +B\left(-\frac{T\sin\left(T\sqrt{4B^{2}+\omega^{2}}\right)}{\sqrt{4B^{2}+\omega^{2}}}+\frac{1-\cos\left(T\sqrt{4B^{2}+\omega^{2}}\right)}{4B^{2}+\omega^{2}}\right)\sz.
\end{aligned}
\end{equation}
And the optimal quantum Fisher information for $\omega$ is
\begin{equation}
\fq[\omega,0]=\frac{4B^{2}T^{2}}{4B^{2}+\omega^{2}}-\frac{8B^{2}T\sin\left(T\sqrt{4B^{2}+\omega^{2}}\right)}{\left(4B^{2}+\omega^{2}\right)^{3/2}}+\frac{8B^{2}\left(1-\cos\left(T\sqrt{4B^{2}+\omega^{2}}\right)\right)}{\left(4B^{2}+\omega^{2}\right)^{2}}.\label{eq:info-omega-no-control}
\end{equation}

\refstepcounter{suppnote}

\section*{Supplementary Note 4. Fisher information for $\boldsymbol{\omega}$
in the presence of Hamiltonian control with $\boldsymbol{\protect\oc}$
near $\boldsymbol{\omega}$\label{note4}}

This Supplementary Note is to obtain the Fisher information for the
rotation frequency $\omega$ of the magnetic field in the presence
of the control Hamiltonian with the control parameter $\oc$ close
to $\omega$.

It has been obtained in the main manuscript that the maximum Fisher
information of estimating $\omega$ with the optimal Hamiltonian control
is
\begin{equation}
\fq[\omega]=B^{2}T^{4},
\end{equation}
and the optimal control Hamiltonian is
\begin{equation}
\hc=B(\cos\omega t\sx+\sin\omega t\sz)-\frac{\omega}{2}\sy.
\end{equation}

However, one generally does not know the exact value of $\omega$,
so the real control that can be applied on the qubit is
\begin{equation}
\hc=B(\cos\oc t\sx+\sin\oc t\sz)-\frac{\oc}{2}\sy,\label{eq:ham-ctrl}
\end{equation}
where $\oc$ is a control parameter that is close to $\omega$, and
the maximum Fisher information is saturated when $\oc=\omega$. In
this case, the total Hamiltonian is
\begin{equation}
\htot=-B(\cos\omega t\sx+\sin\omega t\sz)+B(\cos\oc t\sx+\sin\oc t\sz)-\frac{\oc}{2}\sy.\label{eq:htot-omega}
\end{equation}

A natural question is how the deviation of $\oc$ from $\omega$ influences
the Fisher information. Will a slight deviation of $\oc$ from $\omega$
cause a large drop in Fisher information? This is important to the
design and realization of the Hamiltonian control scheme in practice.

In this Supplementary Note, we derive the Fisher information of estimating
$\omega$ when $\oc$ is deviated from the true value of $\omega$.
We will show how to use this result to design an adaptive control
scheme to approach the $T^{4}$ scaling of the Fisher information
in the next Supplementary Note.

Since the maximum quantum Fisher information of estimating $\omega$
is determined by the gap between the highest and lowest levels of
$\hg[\omega]$, we need to obtain $\hg[\omega]$ first. As $\oc$
is close to $\omega$, we can expand $\hg[\omega]$ with $\oc$ near
$\omega$.

First, according to Eq. (\ref{eq:2}), we have
\begin{equation}
\hg[\omega]\big|_{\oc=\omega}=\int_{0}^{T}\Big[\utd[][t]\po\htot\ut[][t]\Big]_{\oc=\omega}\dt=-\frac{BT^{2}}{2}\sz,
\end{equation}
where $\ut[][t]$ is the unitary evolution operator under $\htot$,
$\ut[][t]|_{\oc=\omega}=\exp(\i\frac{\omega}{2}t\sy)$. Also from
(\ref{eq:2}), we can see that
\begin{equation}
\poc\hg[\omega]\big|_{\oc=\omega}=\int_{0}^{T}\Big[\poc\utd[][t]\po\htot\ut[][t]+\utd[][t]\po\htot\poc\ut[][t]\Big]_{\oc=\omega}\dt,
\end{equation}
where we used $\poc\po\htot=0$. From Eq. (\ref{eq:dgut}), it can
be obtained that
\begin{equation}
\begin{aligned}\poc\ut[][t]\big|_{\oc=\omega}= & -\i\int_{0}^{t}\Big[\ut[\tp][t]\poc\htot[\tp]\ut[0][\tp]\Big]_{\oc=\omega}\dtp\\
= & \frac{t}{2}\sin\frac{\omega t}{2}(-I+\i Bt\sx)+\i\frac{t}{2}\cos\frac{\omega t}{2}(\sy-Bt\sz),
\end{aligned}
\end{equation}
so,
\begin{equation}
\poc\hg[\omega]\big|_{\oc=\omega}=-\frac{BT^{3}}{3}\sx.
\end{equation}

Similarly, according to Eq. (\ref{eq:2}),
\begin{equation}
\begin{aligned}\poct\hg[\omega]\big|_{\oc=\omega}= & \int_{0}^{T}\Big[\poct\utd[][t]\po\htot\ut[][t]+2\poc\utd[][t]\po\htot\poc\ut[][t]\\
 & +\utd[][t]\po\htot\poct\ut[][t]\Big]_{\oc=\omega}\dt,
\end{aligned}
\end{equation}
where we again used $\poc\po\htot=0$. From Eq. (\ref{eq:dgut}),
it can be seen that
\begin{equation}
\begin{aligned}\poct\ut[][t]\big|_{\oc=\omega}= & -\i\int_{0}^{t}\Big[\poc\ut[\tp][t]\poc\htot[\tp]\ut[0][\tp]+\ut[\tp][t]\poct\htot[\tp]\ut[0][\tp]\\
 & +\ut[\tp][t]\poc\htot[\tp]\poc\ut[0][\tp]\Big]_{\oc=\omega}\dtp,
\end{aligned}
\end{equation}
and
\begin{equation}
\poc\ut[\tp][t]\big|_{\oc=\omega}=-\i\int_{\tp}^{t}\Big[\ut[\tpp][t]\poc\htot[\tpp]\ut[\tp][\tpp]\Big]_{\oc=\omega}\d\tpp,
\end{equation}
so it can be obtained that
\begin{equation}
\poct\ut[][t]\big|_{\oc=\omega}=-\frac{1}{4}T^{2}(1+B^{2}T^{2})(\cos\frac{\omega T}{2}I+\i\sin\frac{\omega T}{2}\sy)+\frac{\i}{6}BT^{3}(\cos\frac{\omega T}{2}\sx+\sin\frac{\omega T}{2}\sz),
\end{equation}
thus,
\begin{equation}
\poct\hg[\omega]\big|_{\oc=\omega}=\frac{4B^{2}T^{5}}{15}\sy+\frac{BT^{4}}{4}\sz.
\end{equation}
Therefore, $\hg[\omega]$ can be expanded in the vicinity of $\oc=\omega$
as
\begin{equation}
\hg[\omega]=-\frac{BT^{2}}{2}\sz-\frac{BT^{3}}{3}\sx\do+\left(\frac{4B^{2}T^{5}}{15}\sy+\frac{BT^{4}}{4}\sz\right)\frac{\do^{2}}{2}+O(\do^{3}),\label{eq:approxhomega}
\end{equation}
where $\do=\oc-\omega$.

The eigenvalues of $\hg[\omega]$ in Eq. (\ref{eq:approxhomega})
are
\begin{equation}
\lambda_{\pm}=\pm\frac{BT^{2}}{2}\mp\frac{1}{72}BT^{4}\delta\omega^{2}+O(\do^{4}),
\end{equation}
so, finally, the Fisher information for $\omega$ with $\oc$ near
$\omega$ is
\begin{equation}
\fq[\omega]=B^{2}T^{4}(1-\frac{1}{18}T^{2}\delta\omega^{2}),\label{eq:fisher-expansion}
\end{equation}
where higher order terms of $\do$ have been neglected.

This shows that when $\oc$ is deviated a little from the true value
of $\omega$, the drop in Fisher information is only of the order
$\do^{2}$, and the $T^{4}$ scaling is unaffected, implying the reliability
of the above Hamiltonian control scheme. It lays the foundation for
the feedback control scheme to approach the $T^{4}$ scaling which
will be introduced in the next Supplementary Note.

\refstepcounter{suppnote}

\section*{Supplementary Note 5. Adaptive Hamiltonian control for estimating
$\boldsymbol{\omega}$\label{note5}}

Basing on the result in Supplementary Note 4, we can design schemes
to approach the $T^{4}$ scaling of the Fisher information for estimating
$\omega$ when the exact value of $\omega$ is unknown and an estimate
of $\omega$ is used in the control Hamiltonian.

We first give an overall analysis about why the Hamiltonian control
may increase the Fisher information. In order to apply the Hamiltonian
control scheme, we need an estimate of $\omega$ to be used as the
control parameter $\oc$ in the control Hamiltonian. But if the precision
of the estimate of $\omega$ needed is comparable to the precision
that can be gained, the Hamiltonian control scheme would be senseless.
Fortunately, Eq. (\ref{eq:fisher-expansion}) shows that an approximate
$\omega$ with a precision $\do$ of order $1/T$ is sufficient to
produce an estimate of $\omega$ with a precision of order $T^{-2}$
(i.e., the Fisher information is of order $T^{4}$). This guarantees
that the Fisher information can be increased by the Hamiltonian control
scheme.

The simplest feedback control scheme is just to first obtain an initial
estimate of $\omega$ without any Hamiltonian control, then use it
in the Hamiltonian control (\ref{eq:ham-ctrl}) to produce a high
precision estimate of $\omega$, without any iteration of the scheme.
If the initial estimate of $\omega$ without Hamiltonian control is
sufficiently good, Eq. (\ref{eq:fisher-expansion}) guarantees that
the final estimate of $\omega$ with the Hamiltonian control can attain
the Fisher information of order $T^{4}$.

For the initial estimation of $\omega$, suppose the qubit evolves
for time $T$ without any Hamiltonian control and then it is measured
to estimate $\omega$, and this procedure is repeated for $N$ times.
According to Eq. (\ref{eq:info-omega-no-control}), the variance of
the estimate is
\begin{equation}
\av{\do^{2}}=\frac{4B^{2}+\omega^{2}}{4NB^{2}T^{2}},
\end{equation}
where we assumed $T\gg1/\sqrt{4B^{2}+\omega^{2}}$ to neglect the
lower order terms of $T$ for simplicity. Then if we apply the control
Hamiltonian with this estimate of $\omega$ and let the qubit evolve
for the same time $T$, the Fisher information of the final estimation
will be
\begin{equation}
\fq[\omega]=B^{2}T^{4}\Big(1-\frac{4B^{2}+\omega^{2}}{72NB^{2}}\Big).
\end{equation}
Obviously, if $N\gg\frac{1}{18}(1+\frac{\omega^{2}}{4B^{2}})$, then
\begin{equation}
\fq[\omega]\approx B^{2}T^{4},
\end{equation}
which reaches the $T^{4}$ scaling.

The price for this scheme is that we need additional time $NT$ to
obtain an initial estimate of $\omega$ in the absence of the Hamiltonian
control. But as the additional time $NT$ is linear with $T$, the
$T^{4}$ scaling with the Hamiltonian control can still significantly
outperform the usual $T^{2}$ scaling without the Hamiltonian control.

However, we can also feed the new estimate of $\omega$ from the estimation
with the Hamiltonian control back into the control Hamiltonian so
that a better estimate of $\omega$ can be produced, and this step
can be iterated to further refine the estimate of $\omega$. This
will make the feedback control scheme more efficient.

In the following, we give a detailed analysis about such an adaptive
feedback control scheme, and show a minimum requirement for the precision
of the initial estimation of $\omega$ without Hamiltonian control
to make the feedback control scheme work.

For the sake of generality, we assume the Fisher information of each
measurement in the initial estimation without Hamiltonian control
is $I_{0}$, and the measurement is repeated $N$ times, then the
variance of the initial estimate of $\omega$ is
\begin{equation}
\av{\do^{2}}=\frac{1}{NI_{0}}.
\end{equation}
Now, we apply the Hamiltonian control with this estimate of $\omega$
and choose the evolution time as $T_{1}=\sqrt{I_{0}}$, then the Fisher
information will be
\begin{equation}
I_{1}=B^{2}I_{0}^{2}(1-\frac{1}{18N}).\label{eq:i1}
\end{equation}
We repeat the measurement $N$ times, and use the new estimate of
$\omega$ in the control Hamiltonian. Let the qubit evolve for time
$T_{2}=\sqrt{I_{1}}$, then the Fisher information becomes
\begin{equation}
I_{2}=B^{2}I_{1}^{2}(1-\frac{1}{18N})=B^{6}I_{0}^{4}(1-\frac{1}{18N})^{3}.
\end{equation}
If we iterate this process for $n$ rounds and choose the evolution
time in the $n$-th round to be $T_{n}=\sqrt{I_{n-1}}$, the Fisher
information is
\begin{equation}
I_{n}=B^{2}I_{n-1}^{2}(1-\frac{1}{18N}).
\end{equation}

It can be derived from this iterative relation that
\begin{equation}
T_{n}=I_{0}^{2^{n-2}}[B^{2}(1-\frac{1}{18N})]^{2^{n-2}-\frac{1}{2}},\label{eq:tn}
\end{equation}
and
\begin{equation}
I_{n}=I_{0}^{2^{n}}[B^{2}(1-\frac{1}{18N})]^{2^{n}-1}.
\end{equation}
In terms of $T_{n}$, $I_{n}$ can be rewritten as
\begin{equation}
I_{n}=B^{2}T_{n}^{4}(1-\frac{1}{18N}).
\end{equation}
If $N$ is large, $I_{n}\approx B^{2}T_{n}^{4}$ then. This recovers
the $T^{4}$ scaling of the Fisher information in the presence of
Hamiltonian control.

A question is how many rounds of feedback control it takes to reach
$T_{n}=T$ for a desired time $T$. From Eq. (\ref{eq:tn}), it can
be obtained that
\begin{equation}
n=\left\lceil \log_{2}\frac{\ln\Big(BT\sqrt{1-\frac{1}{18N}}\Big)}{\ln\Big(B^{2}I_{0}(1-\frac{1}{18N})\Big)}\right\rceil +2.\label{eq:nsteps}
\end{equation}
It can be seen that when $T$ is large,
\begin{equation}
n\sim\left\lceil \log_{2}\ln T\right\rceil ,
\end{equation}
which is a double logarithm of $T$, implying $n$ increases extremely
slowly with $T$ and very few rounds of feedback control is sufficient
to reach a large $T$. So this iterative feedback control scheme converges
to the $T^{4}$ scaling very fast.

We also want to remark that there is a lower bound that $I_{0}$ must
satisfy in order to increase the Fisher information by this method.
For Eq. (\ref{eq:i1}), $I_{1}$ is not always larger than $I_{0}$,
and if we want to increase the Fisher information, we need $I_{1}>I_{0}$,
so it requires
\begin{equation}
I_{0}>\frac{1}{B^{2}(1-\frac{1}{18N})}.\label{eq:I0-threshold}
\end{equation}

When $I_{0}$ is above this threshold, $I_{1}$ will also be larger
than $\frac{1}{B^{2}(1-\frac{1}{18N})}$ as the Fisher information
increases. This again makes $I_{2}>I_{1}$. Repeating this step, we
will find that the Fisher information is always increasing, and thus
the allowed $T$ for the $T^{4}$ scaling is also increasing. On the
contrary, if $I_{0}$ does not satisfy the lower bound in Eq. (\ref{eq:I0-threshold}),
$I_{1}$ would be smaller than $I_{0}$, $I_{2}$ would be smaller
than $I_{1}$, and so forth. Therefore, Eq. (\ref{eq:I0-threshold})
is the minimum requirement for the precision of the initial estimation
of $\omega$ without Hamiltonian control in order to make the iterative
feedback control scheme work.

If $T$ is so small that Eq. (\ref{eq:nsteps}) becomes negative,
we just need to set $n=1$ and the $T^{4}$ scaling can be attained.
Note that the $T_{n}$ given in Eq. (\ref{eq:tn}) is just the upper
bound of $T$ that the $T^{4}$ scaling can be achieved (up to a relative
loss of $1/N$) in the $n$-th round, it does not mean that $T$ must
be chosen as $T_{n}$ in that round.

We summarize this iterative feedback control scheme as follows. First,
a rough estimation of $\omega$ should be made in the absence of the
Hamiltonian control with the Fisher information satisfying the threshold
(\ref{eq:I0-threshold}), then the estimate of $\omega$ is used to
apply the Hamiltonian control and get a new estimate of $\omega$.
The new estimate of $\omega$ can be fed back to the control Hamiltonian
to further update the estimate of $\omega$. This procedure is repeated
for $n$ rounds until the desired evolution time $T$ is reached,
and the number of rounds is given by Eq. (\ref{eq:nsteps}). The relative
loss of Fisher information due to the finite number of measurements
is $\frac{1}{18N}$, where $N$ is the number of measurements in each
round of the feedback control scheme, which does not affect the $T^{4}$
scaling.

\refstepcounter{suppnote}

\section*{Supplementary Note 6. Additional control at level crossings of $\boldsymbol{\protect\phgt}$\label{note6}}

In this Supplementary Note, we analyze the effect of the additional
control $\ha$ proposed in the Methods of the main manuscript.

We first prove that under an arbitrary unitary transformation $\vt$
on the states, a Hamiltonian $\ham$ is transformed to
\begin{equation}
\hp=\i(\pt\vt)\vtd+\vt\ham\vtd.\label{eq:5}
\end{equation}
Consider an arbitrary state $\pht$ evolving under $\ham$. It satisfies
the Schr\"odinger equation
\begin{equation}
\i\pt\pht=\ham\pht.
\end{equation}
Now, if we transform the state $\pht$ to $\vpht=\vt\pht$, then $\vpht$
satisfies
\begin{equation}
\begin{aligned}\i\pt\vpht= & \i\pt(\vt\pht)\\
= & \i(\pt\vt)\pht+\i\vt\ppht\\
= & \i(\pt\vt)\vtd\vpht+\vt\ham\vtd\vpht.
\end{aligned}
\end{equation}
Therefore, the Hamiltonian for $\vpht$ is the $\hp$ in Eq. (\ref{eq:5}).

Now we return to our problem. In the presence of the \emph{optimal}
control Hamiltonian, the total Hamiltonian is
\begin{equation}
\htot=\hgt+\hc+\ha,
\end{equation}
where
\begin{equation}
\begin{aligned}\hc & =\sum_{k}\fkt\klm[][k]\blm+\i\sum_{k}\klm[][k]\blm[][][\pt]-\hgt,\\
\ha & =\hkt{}[\etkj[m][n]\klm[][n]\blm[][m]+\etkj[n][m]\klm[][m]\blm[][n]],
\end{aligned}
\end{equation}
and
\begin{equation}
\thk=\int_{0}^{t}\fkt[\tp]\dtp.
\end{equation}

To see the effect of the total Hamiltonian clearly, let us transform
the time-dependent eigenbasis $\{\netk\klm[t][k]\}$ of $\phgt$ to
an arbitrary \emph{time-independent} basis $\{\kk\}$. The transformation
can be written as
\begin{equation}
\vt=\sum_{k}\etk\kk\blm.\label{eq:ut}
\end{equation}
Then the Hamiltonian in the new basis can be obtained from (\ref{eq:5}).

It is straightforward to verify that
\begin{equation}
\i(\pt\vt)\vtd=\i\sum_{kj}\etkj\bklm[][][][\pt]\kk\bj-\sum_{k}\fkt\kk\bj[k].
\end{equation}
and
\begin{equation}
\begin{aligned}\vt(\hgt+\hc)\vtd= & \sum_{k}\fkt\kk\bj[k]+\i\sum_{kj}\etkj\bklm[][][][][\pt]\kk\bj,\\
\vt\ha\vtd= & \hkt\smn,
\end{aligned}
\end{equation}
where $\smn$ is the $\sx$-like transition operator between $\kk[m]$
and $\kk[n]$,
\begin{equation}
\smn=\kk[n]\bj[m]+\kk[m]\bj[n].
\end{equation}
So, by the transformation $\vt$ (\ref{eq:ut}), the total Hamiltonian
$\htot$ becomes
\begin{equation}
\htp=\hkt\smn,\label{eq:new additional Hamiltonian}
\end{equation}
where we have used
\begin{equation}
\bklm[][][][\pt]+\bklm[][][][][\pt]=\pt\bklm=\pt\delta_{kj}=0.
\end{equation}

Eq. (\ref{eq:new additional Hamiltonian}) indicates that the state
$\kk[n]$ will be transformed to $\kk[m]$ in the new basis $\{\kk\}$
by the total Hamiltonian with the additional control, so in the original
basis, $\klm[][n]$ will be transformed to $\klm[][m]$ correspondingly,
which is exactly what we want.

Note that $\htp$ commutes between different $t$, so the evolution
in the new basis $\{\kk\}$ under $\htp$ for a short time interval
$\tk-\frac{1}{2}\sdt\leq t\leq\tk+\frac{1}{2}\sdt$ is
\begin{equation}
\upt=\exp\Big[-\i\int_{\tk-\frac{1}{2}\sdt}^{\tk+\frac{1}{2}\sdt}\hkt[\tp]\dtp\smn\Big]=I\cos\int_{\tk-\frac{1}{2}\sdt}^{\tk+\frac{1}{2}\sdt}\hkt[\tp]\dtp-\i\smn\sin\int_{\tk-\frac{1}{2}\sdt}^{\tk+\frac{1}{2}\sdt}\hkt[\tp]\dtp,\label{eq:7}
\end{equation}
where $\tau$ is the moment that the level crossing occurs. To drive
the system exactly from $\klm[][n]$ to $\klm[][m]$, which corresponds
to driving $\kk[n]$ to $\kk[m]$ in the new basis $\{\kk\}$, we
need
\begin{equation}
\cos\int_{\tk-\frac{1}{2}\sdt}^{\tk+\frac{1}{2}\sdt}\hkt[\tp]\dtp=0,
\end{equation}
thus $\hkt$ must satisfy
\begin{equation}
\int_{\tk-\frac{1}{2}\sdt}^{\tk+\frac{1}{2}\sdt}\hkt[\tp]\dtp=\Big(l+\frac{1}{2}\Big)\pi,\;l\in\mathbb{Z}.
\end{equation}
This proves Eq. (23) in the main text.

A notable point in Eq. (\ref{eq:7}) is that since $\sin(l+\frac{1}{2})\pi=(-1)^{l}$,
$\upt$ will introduce an additional phase $(-1)^{l+1}\i$ to $\klm[][m]$
and $\klm[][n]$. This will affect the relative phase and needs to
be taken into account, when the system is in a superposed state involving
$\klm[][m]$ or $\klm[][n]$.

When there are multiple crossings between the maximum/minimum eigenvalue
and other eigenvalues of $\phgt$ during the whole evolution process
$0\leq t\leq T$, an additional Hamiltonian control $\ha$ is needed
at each level crossing.
\end{document}